\documentclass[twocolumn,notitlepage,superscriptaddress,showkeys,showpacs]{revtex4-1}
\usepackage{graphicx}
\usepackage{dcolumn}
\usepackage{booktabs,bm,color,braket}
\usepackage{amsmath}
\usepackage{color}
\usepackage[dvipsnames]{xcolor}
\usepackage[hidelinks,colorlinks,linkcolor=blue,citecolor=blue,urlcolor=blue]{hyperref}

\usepackage{pdfpages}
\usepackage{etoolbox} 
\makeatletter
\patchcmd{\@outputpage@head}{\@ifx{\LS@rot\@undefined}{}{\LS@rot}}{}{}{}
\makeatother

\begin{document}

\title{An electron-counting rule to determine the interlayer magnetic coupling of the van der Waals materials}
\author{Jiewen Xiao}
\affiliation{School of Materials Science and Engineering, Beihang University, Beijing 100191, P. R. China.}
\author{Binghai Yan}\email{binghai.yan@weizmann.ac.il}
\affiliation{Department of Condensed Matter Physics, Weizmann Institute of Science, Rehovot 7610001, Israel}

\begin{abstract}
In layered magnetic materials, the magnetic coupling between neighboring van der Waals layers is challenging to understand and anticipate, although the exchange interaction inside a layer can be well rationalized for example by the superexchange mechanism. In this work, we elucidate the interlayer exchange mechanism and propose an electron-counting rule to determine the interlayer magnetic order between van der Waals layers, based on counting the $d$-orbital occupation ($d^n$, where $n$ is the number of $d$-electrons at the magnetic cation). With this rule, we classify magnetic monolayers into two groups, type-I ($n<5$) and type-II ($n\geq5$), and derive three types of interlayer magnetic coupling for both insulators and metals. The coupling between two type-II layers prefers the antiferromagnetic (AFM) order, while type-I and type-II interface favors the ferromagnetic (FM) way. However, for two type-I layers, they display a competition between FM and AFM orders and even lead to the stacking dependent magnetism. Additionally, metallic layers can also be incorporated into this rule with a minor correction from the free carrier hopping. Therefore, this rule provides a simple guidance to understand the interlayer exchange and further design van der Waals junctions with desired magnetic orders.
\end{abstract}
\maketitle

\section{Introduction}
Magnetic van der Waals materials have received considerable attention, inspired by the successful exfoliation of magnetic monolayers and few layers in experiments, such as semiconducting CrI$_3$ and metallic Fe$_3$GeTe$_2$ \cite{huang2017layer,gong2017discovery,deng2018gate,fei2018two,bonilla2018strong,burch2018magnetism,gong2019two,gibertini2019magnetic}. 
Besides intriguing 2D magnetism, these layers can further form  magnetic multiple layers and heterostructures as novel spintronic devices \cite{Jiang2018,song2018giant,klein2018probing,kim2018one} and topological materials \cite{deng2020quantum,gong2019experimental}.
Inside the van der Waals layer, most materials are ferromagnetic (FM) and can be understood by the conventional superexchange \cite{goodenough1963magnetism} and itinerant exchange \cite{coronado1996exchange} mechanisms. In contrast, the interlayer magnetic coupling across the van der Waals gap is less explored, although it is essential to design magnetic junctions.

The interlayer interaction is found to be subtle. For example, in the representative material CrI$_3$, the interlayer magnetic order, FM or antiferromagnetic (AFM), depends sensitively on the stacking order and the external pressure \cite{sivadas2018stacking,Song2019,jiang2019stacking}.
The interlayer exchange is believed to be weak because of the van der Waals gap, usually leading to low magnetic ordering temperature, e.g., 61 K for the FM CrI$_3$ \cite{McGuire2015} and 24 K for the AFM MnBi$_2$Te$_4$ \cite{otrokov2019prediction}. Surprisingly, it is found very recently that the CrI$_3$ monolayer couples to the MnBi$_2$Te$_4$ layer in a FM way with the large exchange energy of 40 meV \cite{fu2019exchange}.
As separated by the van der Waals gap, the interlayer magnetic exchange originates from indirect exchange pathways, referred to the super-superexchange effect \cite{sivadas2018stacking}.
This is different from the conventional superexchange where single anion serves as an intermediate to bridge two magnetic cations \cite{goodenough1963magnetism}.
It is intriguing but challenging to clarify the microscopic mechanism of interlayer magnetic coupling and provide a generic understanding that is applicable to a wide range of van der Waals magnets. 

In the present work, we aim to build up a general rule to determine interlayer magnetic coupling for van der Waals materials. This rule defines two basic exchange pathways that cross the van der Waals gap (see Figure 1): The AFM exchange between two occupied $d$ orbitals and the FM exchange between one occupied and the other empty $d$ orbital. Only through counting the occupation of $d$ orbitals with different exchange effects, without requiring sophisticated calculations, 
interlayer magnetic order can be anticipated with the satisfactory accuracy for both semiconducting and metallic materials. With this rule, monolayers are classified into type-I ($d^n$, $n<5$) and type-II ($d^n$, $n\geq5$), where $n$ is the $d$-orbital occupation number of the magnetic cation. Subsequently, three types of bilayer-interfaces are identified. Bilayers as type I-II, II-II and I-I display FM, AFM and competing magnetic orders, respectively. In addition, for metallic bilayers, the extra itinerant exchange effect from the interlayer free carrier hopping requires a minor correction to the above rule. Our proposed rule is further verified by first-principles and model Hamiltonian computations. By revealing the interlayer magnetic exchange, our work serves as a simple guidance for the experiment and theory on layered magnetic structures. 

\section{Computational Methods}
Magnetic bilayers, composed of MX$_2$ (M = V, Cr, Mn; X = S, Se) \cite{guo2017modulation,o2018room,wang2018layer,bonilla2018strong}, MY$_2$ (M = Mn, Fe, Co, Ni; Y = Cl, Br) \cite{botana2019electronic}, MI$_3$ (M = V, Cr) \cite{huang2017layer,tian2019ferromagnetic}, and CrGeTe$_3$ \cite{gong2017discovery} monolayers, are investigated. They all display the intralayer FM order except MnY$_2$ (Y = Cl/Br) and CrS$_2$ monolayer phase with the stripy AFM order. Van der Waals magnets considered here include 1-T phase, CrI$_3$ phase and CrGeTe$_3$ phase, and their coordination environments all belong to the distorted octahedral field (Figure S1 in Supplementary Materials). Therefore, the $d$-orbital crystal field is of the $t_{2g}$-$e_{g}$ type. Regarding the experimental progress, we restrict our discussions to materials with $3d$ transitional elements and the colinear intralayer FM coupling in this work.
Therefore, the strong $3d$ onsite Coulomb repulsion ($U$) is usually much larger than the crystal field splitting ($\Delta$). Thus, magnetic cation is assumed to be high spin state while low spin state can also be incorporated as discussed in the following text. 

First-principles calculations have been performed in the framework of density functional theory using the Vienna Ab initio Simulation Package (VASP) \cite{kresse1996efficiency,kresse1996efficient}. Perder-Burke-Ernzerhof (PBE) formulation was applied to describe the exchange-correlation under the generalized gradient approximation (GGA) \cite{perdew1996generalized}. Van der Waals corrections are included by the DFT+D3 method \cite{grimme2010consistent} in the bilayer structural optimization. Considering the localized nature of 3$d$ electrons for transition metals, the GGA+U method was adopted \cite{anisimov1997first}, where the effective $U-J$ value was set to 3 eV, a typical value for $3d$ transitional elements. Spin-orbit coupling (SOC) is not considered here, which is generally weak compared to the exchange coupling in 3$d$ systems.


\section{Results and Discussions}

\subsection{The rule for the interlayer exchange coupling}
\begin{figure}[t]
    \centering
    \includegraphics{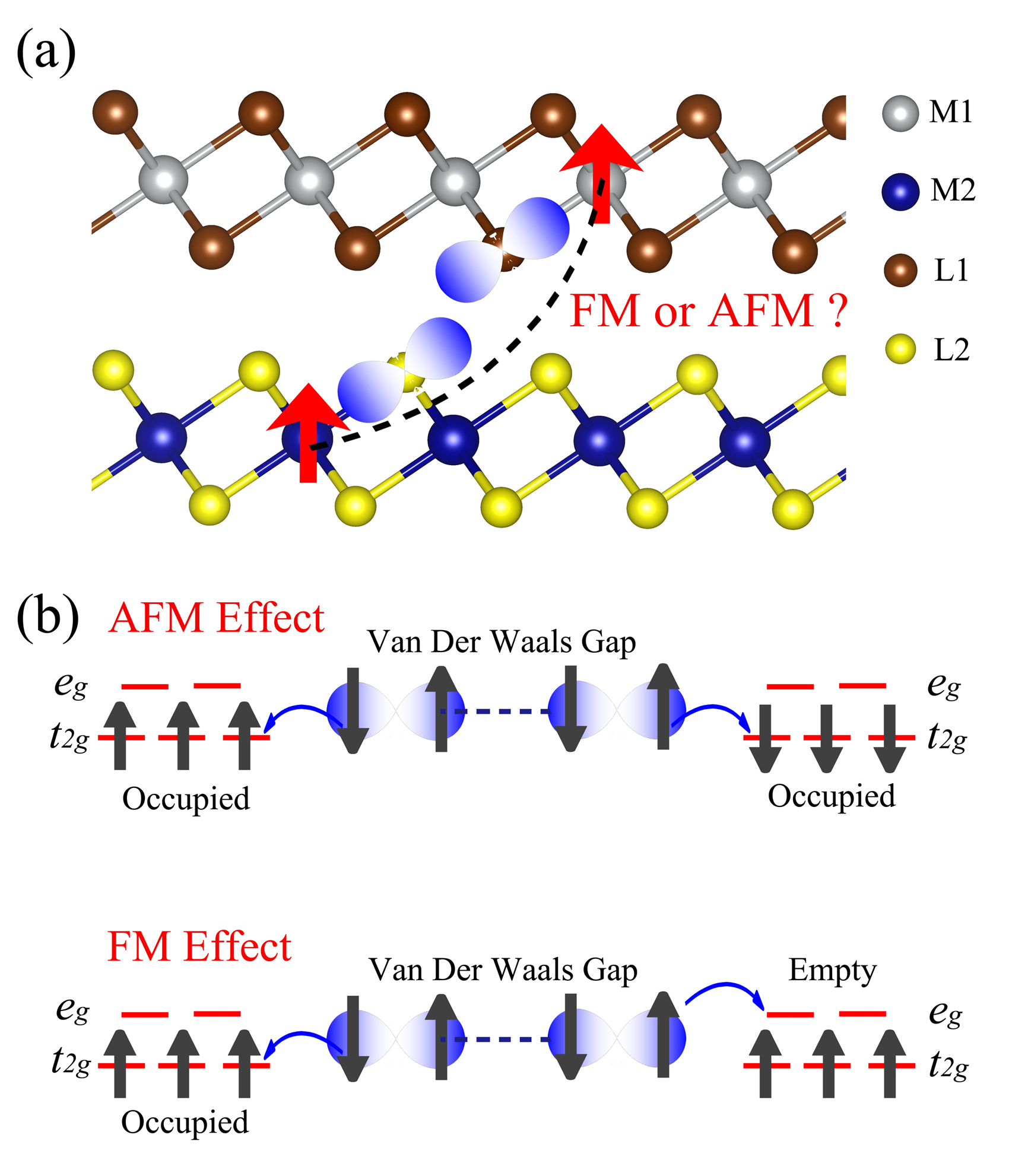}
    \caption{\textbf{Schematics of the interlayer exchange coupling. }
    a. Magnetic exchange coupling between two van der Waals layers via ligand atoms. M1, M2 and L1, L2 denote magnetic cations and ligands, respectively. Red arrows represent spins on magnetic cations. The blue dumbbell shapes represent the $p$ orbitals of ligands. b. Two basic exchange pathways with the AFM and FM exchange effects.}
    \label{Figure 1}
\end{figure}

\begin{table*}
\caption{Interlayer magnetic orders for insulating bilayers with varied electronic configurations.}
\label{table 1}
\begin{tabular}{l}
\includegraphics{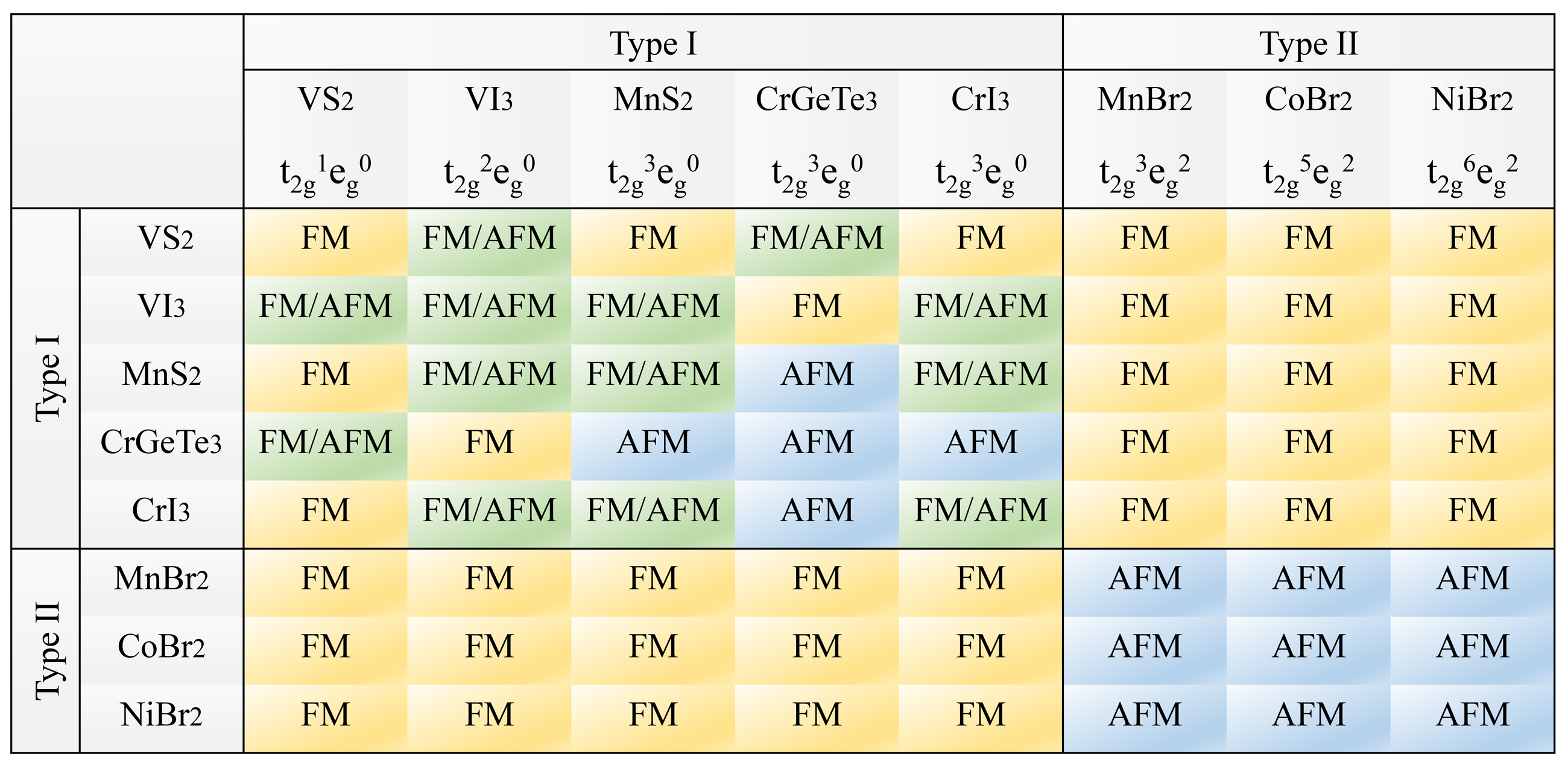}
\end{tabular}
\end{table*}

Interlayer magnetic coupling between van der Waals layers originates from the indirect exchange interaction, where $p$ orbitals in adjacent layers intermediates intralayer $d$ orbitals, as presented in Figure 1(a). Here we concentrate on interlayer magnetic coupling and intralayer magnetic order is assumed to be FM. To understand the interlayer interaction, we classify two basic exchange pathways, as shown in Figure 1(b): (i) AFM exchange interaction between two occupied $d$ orbitals, (ii) FM exchange interaction between one occupied and the other empty $d$ orbital. 

Two basic exchange pathways can be rationalized by two elementary process: intralayer $d$-$p$ hopping and interlayer $p$-$p$ interaction. The former one is stronger and spin selective based on the occupation of $d$ orbitals, as shown in Figure 1. However, the latter one is subtle. For the weak interlayer $p$-$p$ interaction, we postulate that it favors the anti-parallel alignment of $p$ electrons that are separated by van der Waals gap, since this alignment allows interlayer electron hopping and has the stronger kinetic energy contribution.

To determine the interlayer magnetic order, we need to count all the basic exchange pathways based on the $d$ orbital occupation and evaluate the total exchange interaction. 
To further proceed, we categorize two kinds of monolayers as type-I and type-II, with electronic states as $d^n$ ($n<5$) and $d^n$ ($n\geq5$). Since van der Waals magnets considered here belong to the distorted octahedral ligand filed with the high spin state (Figure S1 in Supplementary Materials), electronic configurations can be denoted as type-I $t_{2g}^xe_g^y$ ($x+y<5$) and type-II $t_{2g}^{x}e_g^{y}$ ($x+y\geq5$). Compared to type-I monolayer, all $d$ orbitals in type-II layer are occupied with no empty $d$ orbitals available. Therefore, two types of monolayers construct three types of bilayers: type I-I, II-II and I-II. With this classification, we can intuitively predict that the type II-II bilayer exhibits the interlayer AFM order, since only occupied to occupied exchange, i.e. the AFM coupling, pathways exist. In contrast, both the type I-I and type I-II bilayers display competing FM and AFM orders, since two types of pathways (occupied to occupied, and occupied to empty) coexist. While for type I-II, we further find that the FM coupling is usually more favorable because of the orbital orientation and large onsite $U$, as discussed in the following text.

\subsection{Insulating magnetic layers}

To verify the above scenario, we first investigated insulating bilayers with varied electronic configurations and stacking orders. First-principles results are presented in Table I and the detailed energy differences between interlayer FM and AFM order against stacking orders are shown in Section II in Supplementary Materials. 
Firstly, for type I-I bilayer composed of MS$_2$ (M = V, Mn), VI$_3$, CrI$_3$, and CrGeTe$_3$ monolayers, as what we predicted, they show competing magnetic orders, which originate from both the existence of FM and AFM exchange pathways. For instance, bilayer phase of VS$_2$ exhibits interlayer FM order, in accordance with experiments and their corresponding bulk phases \cite{guo2017modulation}. CrGeTe$_3$-MnS$_2$ heterostructure and CrGeTe$_3$ bilayers display the interlayer AFM order. However, most bilayers and heterostructures display the stacking-dependent magnetism, as denoted by “FM/AFM” in the green box of Table I. Such a stacking-dependent magnetic order is a manifestation of the competing exchange pathways.

For type II-II bilayers and heterostructures composed of MX$_2$ (M = Mn, Co, Ni, X = Cl/Br), they always favor the AFM coupling since only the AFM exchange pathway between occupied $d$ orbitals is available. This type of bilayer also incorporates few layer phases of MnBi$_2$Te$_4$, since the electronic state of magnetic cation Mn$^{2+}$ is $t_{2g}^3e_g^2$. For type I-II bilayers, all tested heterostructures exhibit FM order, for instance CrI$_3$-MBr$_2$ (M = Mn, Co, Ni). Therefore, the FM exchange pathway is always more favorable than the AFM one. Our previous work also identified the interlayer FM coupling for CrI$_3$-MnBi$_2$Te$_4$ heterostructure \cite{fu2019exchange}, which can be incorporated into this type I-II system.

\begin{figure*}[t]
    \centering
    \includegraphics{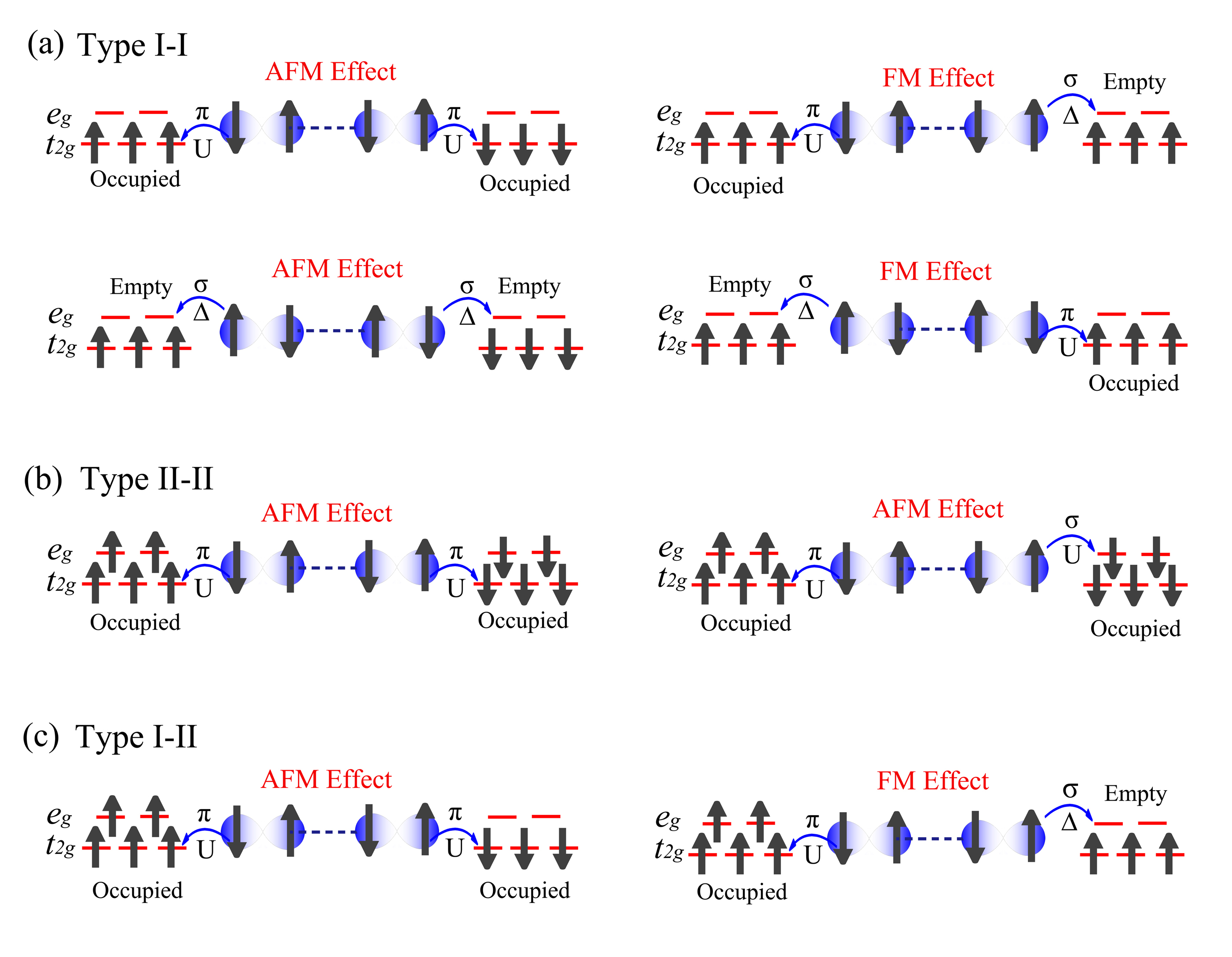}
    \caption{Exchange pathways for a. type I-I, b. type II-II and c. type I-II bilayers. $\Delta$ and $U$ represent the $t_{2g}$-$e_g$ crystal field splitting and onsite Hubbard energy, respectively. $\pi$ and $\sigma$ represent the $d$-$p$ atomic bonding type.}
    \label{Figure 2}
\end{figure*}

To understand Table I in detail, we elaborated exchange pathways for different bilayers based on $d$-orbital occupation. Figure 2(a) presents a typical electronic configuration as $t_{2g}^3e_g^0$-$t_{2g}^3e_g^0$ for type I-I bilayer. On the one hand, the AFM coupling results from two exchange pathways: (1) magnetic interaction between two occupied $t_{2g}$ orbitals ($t_{2g}$-$p$-$p$-$t_{2g}$); (2) magnetic interaction between two empty $e_g$ orbitals ($e_g$-$p$-$p$-$e_g$). On the other hand, there also exists two FM exchange pathways between one empty $e_g$ and one occupied $t_{2g}$ orbital in different layers ($e_g$-$p$-$p$-$t_{2g}$ and $t_{2g}$-$p$-$p$-$e_g$). In Figure 2(a), the basic $d$-$p$ hopping process further indicates that the $t_{2g}$-$p$ hopping involves the onsite $U$ energy with the $\pi$-bonding between $t_{2g}$ and $p$ orbitals. On the other hand, the $e_g$-$p$ hopping is related to the crystal field splitting energy $\Delta$ with the $\sigma$-bonding. Because of $U>\Delta$ and the stronger $\sigma$-bonding compared to the $\pi$-bonding, $e_g$-$p$ hopping is generally more favorable than the $t_{2g}$-$p$ process.

However, the stronger $e_g$-$p$ hopping exists in both AFM $e_g$-$p$-$p$-$e_g$ and FM $e_g$-$p$-$p$-$t_{2g}$ and $t_{2g}$-$p$-$p$-$e_g$, and their combined effect is the uncertain interlayer magnetic order. While for the weak interlayer $p$-$p$ interaction, it can be tuned by stacking orders and affects AFM $e_g$-$p$-$p$-$e_g$ and $t_{2g}$-$p$-$p$-$t_{2g}$ and FM $e_g$-$p$-$p$-$t_{2g}$ and $t_{2g}$-$p$-$p$-$e_g$ to different extent, thus leading to the stacking dependent magnetism. For other electronic configurations of type I-I bilayer, both the empty and occupied $d$ orbitals generally exist, whose interaction scenario is similar to the $t_{2g}^3e_g^0$-$t_{2g}^3e_g^0$ system, as presented in Section III in Supplementary Materials. 

However, the situation is simpler for the type I-II interface.
In Figure 2(c), AFM effect is from the interaction between two occupied $t_{2g}$ orbitals in type I and type II layer ($t_{2g}$-$p$-$p$-$t_{2g}$), which competes with the FM effect produced by the interaction between one empty $e_g$ in type I and one occupied $t_{2g}$ in type II ($e_g$-$p$-$p$-$t_{2g}$). 
Comparing AFM and FM pathways, the $\sigma$-type $e_g$-$p$ hopping with the $\Delta$ gap is much stronger than the corresponding $\pi$ type $t_{2g}$-$p$ hopping with the $U$ gap. Thus, we obtain the FM order in the end. Finally, for type II-II bilayer in Figure 2(b), it is not surprising that interlayer interaction always favors AFM order, since only AFM pathways exist.

To summarize, type I-I, type II-II and type I-II bilayers exhibit competing, AFM and FM orders. Generally, when multiple $d$ orbitals exist, through listing exchange pathways based on the $d$ orbital occupation, we can understand different exchange effects and further determine the interlayer magnetic order by considering their competitions. This procedure is also applicable to the low spin state of magnetic cations once their electronic configurations are clarified.

\subsection{Metallic magnetic layers}

The above interlayer exchange coupling applies to both insulating and metallic systems. For metallic bilayers, however, the additional interlayer exchange from itinerant carriers modifies the interaction. To understand this correction, bilayers composed of intrinsic metallic monolayers, including VSe$_2$, CrS$_2$, MnSe$_2$, FeCl$_2$ and FeBr$_2$, are investigated. The former three monolayers belong to type I while FeCl$_2$ and FeBr$_2$ belong to type II. First-principles results are presented in Table II. It shows that the interlayer magnetic order for type I-I, type II-II and type I-II bilayer basically follows the above scenario. But some deviations indeed appear: type I-I bilayers mainly adopt interlayer FM order; some bilayers in type II-II and type I-II region even exhibit stacking dependent magnetism, like type II-II FeCl$_2$-FeBr$_2$ and type I-II VSe$_2$-FeCl$_2$. 

\begin{table}
\caption{Interlayer magnetic orders for metallic bilayers with varied electronic configurations.}
\label{table 2}
\begin{tabular}{l}
\includegraphics{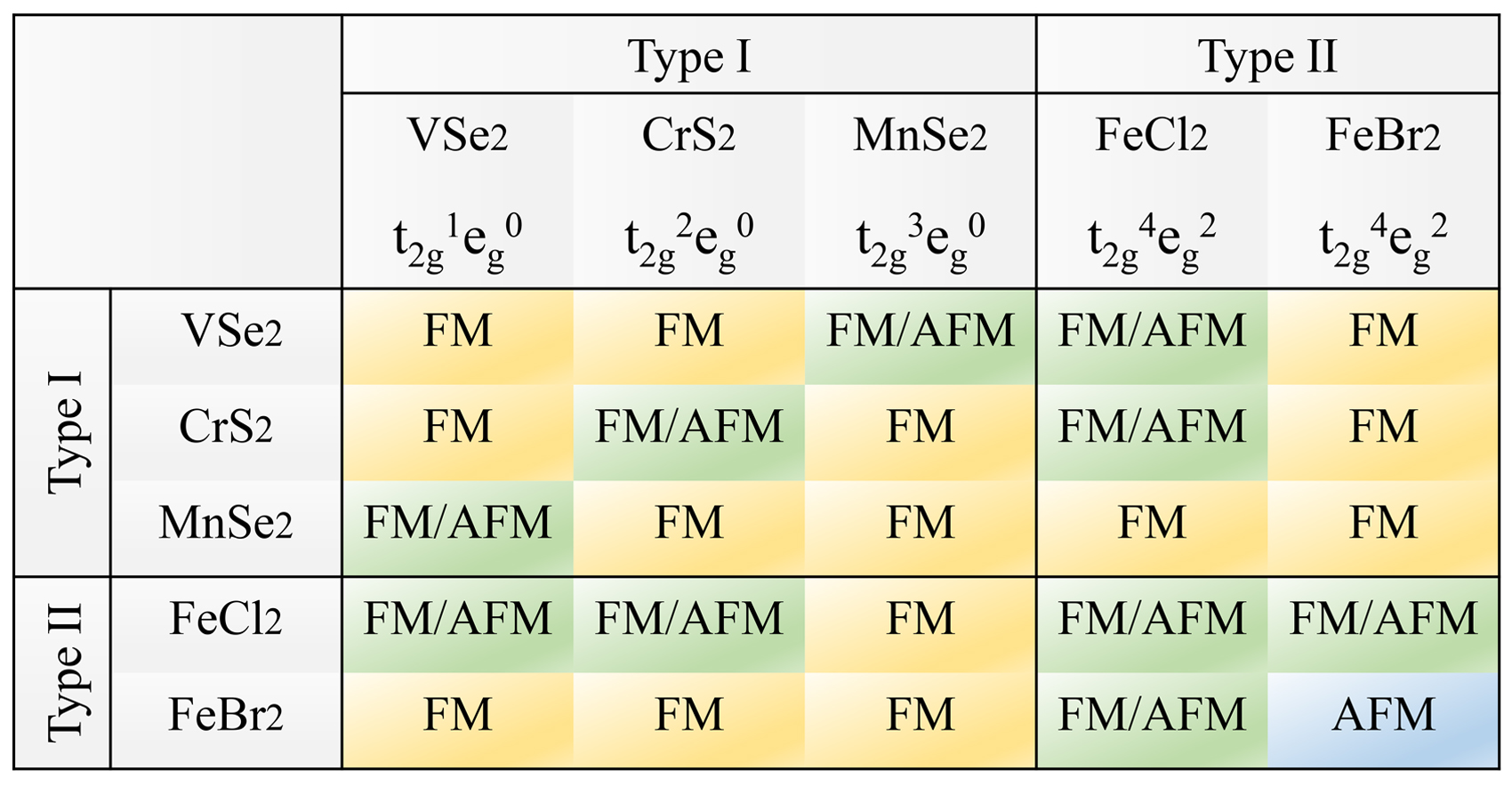}
\end{tabular}
\end{table}
\begin{figure*}[t]
    \centering
    \includegraphics{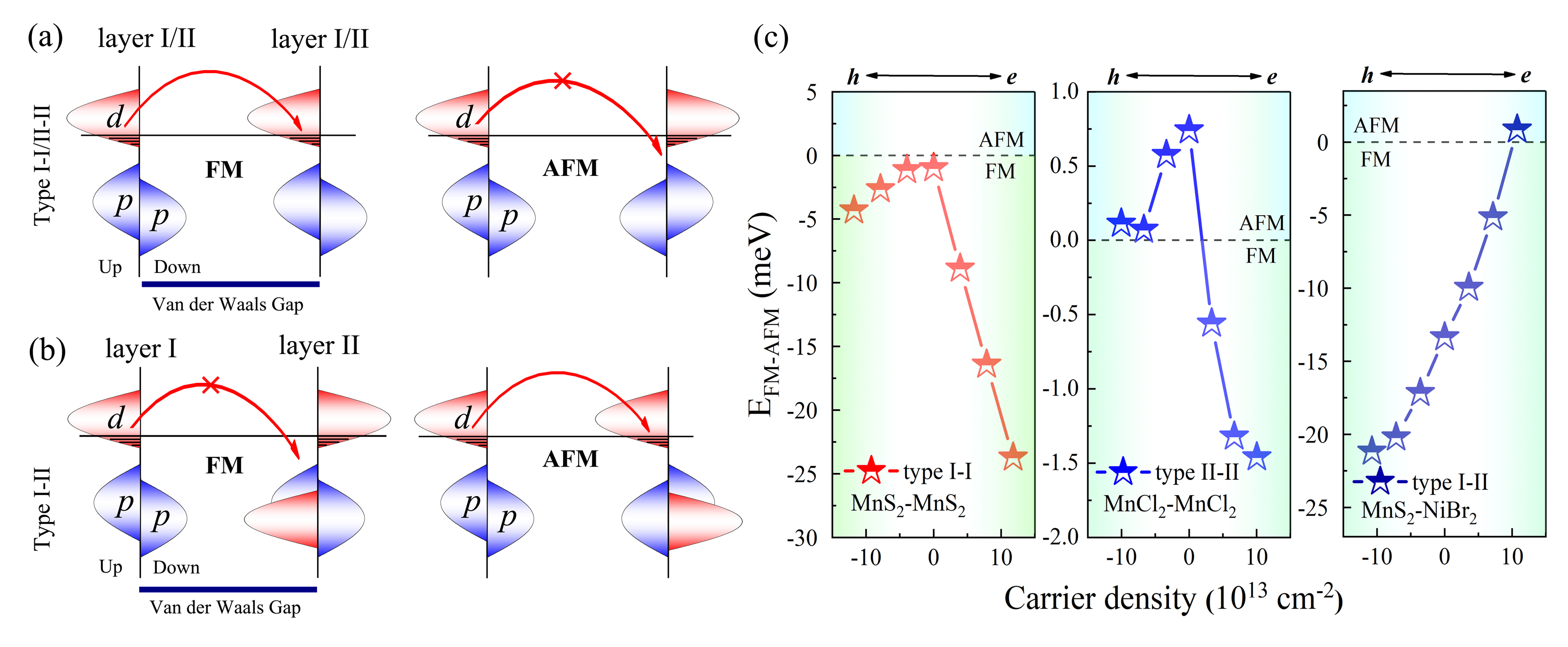}
    \caption{Schematic representation of the interlayer free carrier hopping for a. type I-I and type II-II bilayers and b. type I-II bilayer, respectively. c. Evolution of E$_{FM-AFM}$ (the energy differences between interlayer FM and AFM order) against carrier density for type I-I and type II-II, type I-II bilayers, where the positive/negative carrier density denotes electron/hole doping.}
    \label{Figure 3}
\end{figure*}

To understand this behavior, the interlayer exchange from intralayer itinerant carriers need to be incorporated. Similar to the itinerant magnetism \cite{coronado1996exchange}, the spin configuration that allows the interlayer free carrier hopping is favored, since it contributes to the kinetic energy. Figure 3(a) presents the typical density of states for FM metals. It shows that interlayer carrier hopping is allowed for FM state but impeded for AFM order, since, in AFM order, conducting carriers in the top layer can only hop into gapped states in the lower layer and vice versa. Therefore, free carriers bring the additional FM exchange effect, and this is applicable to both type I-I and type II-II bilayers. For the former, the FM coupling is enhanced. For the latter, the itinerant FM interaction induces the competition with the AFM coupling, thus resulting in the stacking dependent magnetism in type II-II region, as seen in Table II. In addition, we also investigated the itinerant Fe$_3$GeTe$_2$ magnetic bilayer. Fe cation exhibits the $d^8$ configuration and belongs to the type II-II coupling in our classification. We note that Fe$_3$GeTe$_2$ is crystallized into a different structure from the 1T phase and does not exhibit the $t_{2g}$-$e_g$ type crystal field splitting. We found that the bilayer is FM in the ordinary stacking, consistent with recent experiments \cite{fei2018two,deng2018gate}. But we also found that its interlayer coupling depends on the stacking order due to the competition between the AFM exchange and the itinerant FM interaction.

However, type I-II bilayer possesses special electronic configuration, where itinerant electrons in different layers have different spin components: spin up electrons in type I and spin down electrons in type II, as shown in Figure 3(b). Therefore, for AFM order, conducting electrons in type I and type II can hop into each other’s partially occupied states rather than gapped states in the less favorable FM order. As a result, free carriers in type I-II bilayer bring AFM exchange effect that competes with the dominant FM exchange, thus leading to the stacking dependent magnetism in type I-II VSe$_2$-FeCl$_2$ and CrS$_2$-FeCl$_2$. 

Besides intrinsic metallic phases, carrier doped insulating bilayers also display the itinerant exchange effect. As illustrated in Figure 3(c) and S5, for both electron and hole doping, type I-I MnS$_2$-MnS$_2$ and CrI$_3$-CrI$_3$ and type II-II MnCl$_2$-MnCl$_2$ and NiCl$_2$-NiCl$_2$ all manifest the enhanced FM or weakened AFM coupling, with the increasing concentration of free carriers. It can be understood that, FM effect introduced by free carriers gradually modifies the insulating exchange. Interestingly, the phase transition from AFM to FM for CrI$_3$-CrI$_3$, MnCl$_2$-MnCl$_2$, and NiCl$_2$-NiCl$_2$ can even be observed with the increasing of the doping concentration.
This is further consistent with the experimental electrostatic doping control of 2D magnetism in CrI$_3$ bilayer \cite{jiang2018controlling}.
Generally, for type I-I and type II-II bilayers with the intrinsic AFM order, electrostatic doping can introduce FM effect and is a general strategy to effectively tune the phase transition from AFM to FM order. While for type I-II bilayer, we have tested MnS$_2$-NiBr$_2$ and MnS$_2$-MnBr$_2$ heterostructure and results in Figure 3(c) and Figure S5 show that electron doping weakens the intrinsic FM coupling since free carriers in type I and type II layer possess different spin components, as predicted. On the other hand, due to the special alignment of electronic states (Section IV in Supplementary Materials), doped holes possess the same spin components, and still lead to the FM exchange effect. 

To summarize, metallic bilayers basically follow the prediction of the insulating interlayer exchange. But the additional itinerant FM effect for type I-I and type II-II and itinerant AFM effect for type I-II need to be considered to understand some deviations. Even for insulating layers, carrier doping can display the similar itinerant exchange interactions. 

\subsection{Model Hamiltonian for interlayer magnetic coupling}
\begin{figure*}[t]
    \centering
    \includegraphics{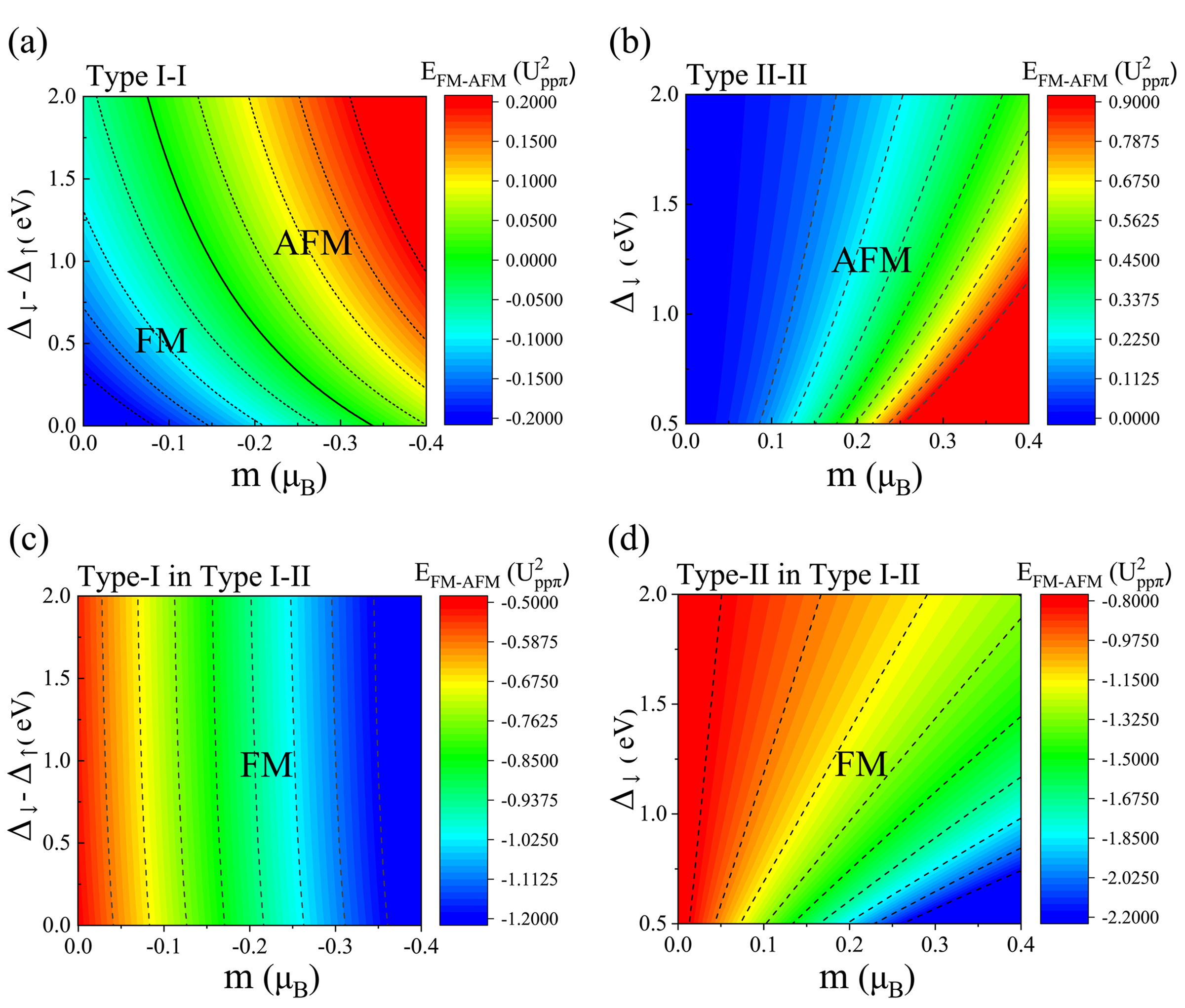}
    \caption{The contour map of E$_{FM-AFM}$ (the energy differences between interlayer FM and AFM order) against magnetic moment on ligands and charge transfer gap for a. type I-I, b. type II-II, and c, d. type I-II bilayers, respectively. $U_{pp\pi}$ is the Slater-Koster hopping parameter for the interlayer $p$-$p$ interaction (See Section V in Supplementary Materials).}
    \label{Figure 4}
\end{figure*}
To verify the basic interlayer exchange in a more rigid way, we construct a phenomenological model Hamiltonian:
\begin{equation}
\begin{split}
      H=\sum_{i,l,u,\sigma}\epsilon_{lu\sigma}c_{ilu\sigma}^{\dagger}c_{ilu\sigma}-t\sum_{ij,l,uv,\sigma}(c_{ilu\sigma}^{\dagger}c_{jlv\sigma}+H.C.)\\+U\sum_{i,l,u}n_{ilu\uparrow}n_{ilu\downarrow}-t_l\sum_{i,l,u,\sigma}(c_{ilu\sigma}^{\dagger}c_{i(1-l)u\sigma}+H.C.)
\end{split}
\end{equation}
Four subsequent terms correspond to onsite energy, kinetic energy of intralayer electron hopping ($t$), onsite Coulomb repulsion ($U$) and kinetic energy of interlayer electron hopping ($t_l$). Here $c_{ilu\sigma}^\dagger$ ($c_{ilu\sigma}$) is the electron creation (annihilation) operator for orbital $u$ at site $i$ and layer $l$ with spin $\sigma$. $l$ has the value 0 and 1, denoting the lower and upper layers. $\epsilon_{lu\sigma}$ is the onsite energy for orbital $u$ ($u = d, p$) with spin $\sigma$ at layer $l$. 
Detailed analytic expressions and approximations are presented in Section V in Supplementary Materials. 

Results in Figure 4 show the phase diagram against induced magnetic moment on ligands $m$ ($\mu_B$) and charge transfer gap $\Delta_{\sigma}$ = $E_{d\sigma}-E_{p\sigma}$ (eV) for type I-I, type II-II and type I-II layers, respectively. It is worth to note that, due to the spin selective $d$-$p$ electron hopping, ligand atoms have negative and positive magnetic moment for type-I and type-II layers. And the stronger $|m|$ reflects the larger the hybridization differences between spin up and spin down $d$-$p$ orbitals. Therefore, ligand magnetic moment $m$ is used as an parameter in the phase diagram, to reflect the intralayer exchange splitting. While analytical expression of $m$ with regard to the $d$-$p$ hopping integral and onsite energies is presented in Supplementary Materials. 

As shown in Figure 4(a), FM and AFM order both exist in type I-I bilayer and the phase transition can be observed. Furthermore, reducing both the $\Delta_{\downarrow}$-$\Delta_{\uparrow}$ and $|m|$ benefits the FM coupling. It suggests that the weakened intralayer exchange splitting favors FM $e_g$-$p$-$p$-$t_{2g}$ and $t_{2g}$-$p$-$p$-$e_g$ exchange pathways. In other words, the reduced differences between $\sigma$ type $e_g$-$p$ hopping with $\Delta$ gap and $\pi$ type $t_{2g}$-$p$ hopping with $U$ gap favors the FM coupling. 
For type II-II bilayer, Figure 4(b) shows that AFM coupling is always favored, but increasing $\Delta_{\downarrow}$ and decreasing ligand polarization can repress the interlayer AFM strength. 
Finally, for type I-II bilayer, FM order is also robust against intralayer parameters. And increasing the ligand polarization $|m|$ in both type-I and type-II layers can further enhance the FM exchange. 
Therefore, the contour map constructed from model Hamiltonian is consistent with out rule, and can also be utilized to modulate the interlayer coupling strength. For instance, the biaxial strain effect is explored in Section V in Supplementary Materials, which can effectively tune the ligand polarization and thus tailor the exchange coupling. 

\section{Conclusion}
To conclude, we proposed a simple electron-counting rule to determine the interlayer magnetic order between van der Waals layers based on the $d$-orbital occupation. Through elaborating exchange pathways, the general competing, AFM and FM interlayer magnetic orders for type I-I, type II-II and type I-II bilayers are predicted and verified by first-principles and model Hamiltonian calculations. In addition, for metallic bilayers, the exchange correction by free carriers is also revealed. 
Our work clarifies the interlayer exchange mechanism and provides guiding principles to design and tailor 2D magnetic materials.

\section{Acknowledgment}
We thank helpful discussions with Huixia Fu, Chaoxing Liu, Ella Orion Lachman, Di Xiao, Jiun-Haw Chu, and Kin Fai Mak. 
B.Y. acknowledges the financial support by the Willner Family Leadership Institute for the Weizmann Institute of Science,
the Benoziyo Endowment Fund for the Advancement of Science, 
Ruth and Herman Albert Scholars Program for New Scientists.
%

\clearpage


\section*{Supplementary material}

# An electron-counting rule to determine the interlayer magnetic coupling of the van der Waals materials

Jiewen Xiao[1] and Binghai Yan[2, *]

[1]*School of Materials Science and Engineering, Beihang University, Beijing 100191, P. R. China.*
[2]*Department of Condensed Matter Physics, Weizmann Institute of Science, Rehovot 7600001, Israel*

## I. GEOMETRIC STRUCTURES OF MAGNETIC MONOLAYERS

Magnetic monolayers, including $MX_2$ (M = V, Cr, Mn; X = S, Se), $MY_2$ (M = Mn, Fe, Co, Ni; Y = Cl, Br), $MI_3$ (M = V, Cr), and $CrGeTe_3$ can be classified into 1-T phase, $CrI_3$ phase, $CrGeTe_3$ phase. Their coordination environments all belong to the distorted octahedral field, as shown in Figure S1.

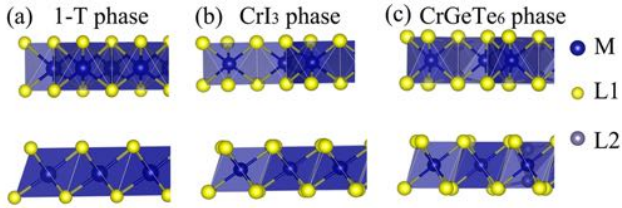

FIG. S1. Distorted octahedral field for a. 1-T phase, b. $CrI_3$ phase and c. $CrGeTe_3$ phase, where M and L1, L2 represent magnetic cations, ligands and Ge atom in $CrGeTe_3$.

## II. INTERLAYER COUPLING STRENGTH AGAINST STACKING ORDERS

For table I and II in the main text, we further present the contour map of the energy difference between interlayer FM and AFM order ($E_{FM-AFM}$) against stacking orders. The origin $(0,0)$ in the contour map denotes the zero lateral shift between top and lower layers. The corresponding $(0,0)$ stacking geometries for different bilayers are presented in Figure S2 and S3. For the general point $(x, y)$ in the contour map, it represents the lateral shift of the top layer by $xa + yb$ from the origin, where $a$ and $b$ are crystal parameters of the bilayer superlattice. All the contour maps for insulating type I-I (Figure S11-S14), type I-II (Figure S15-S18) and type I-II bilayers (Figure S19-S20), and metallic type I-I (Figure S21), type I-II (Figure S22) and type II-II (Figure S23) are presented at the end of the supplementary materials.

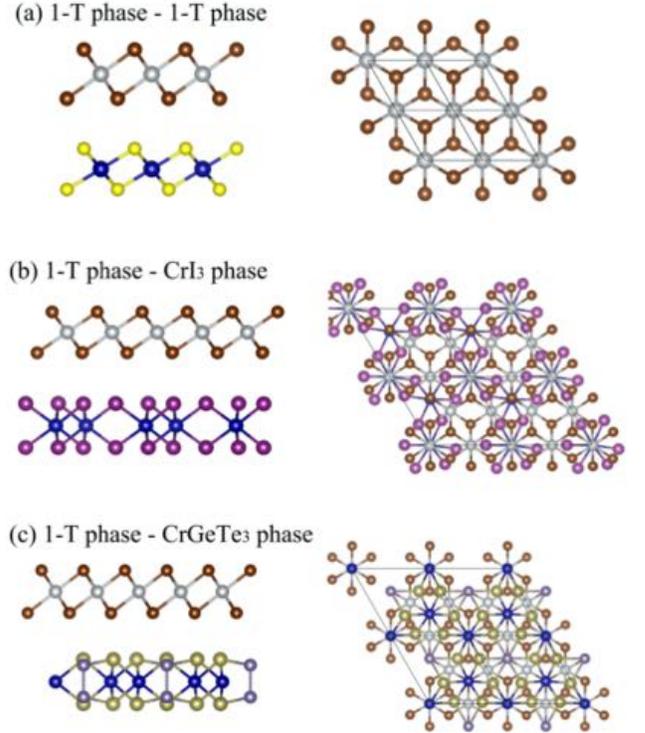

FIG. S2. $(0, 0)$ stacking geometries for a. 1-T phase - 1-T phase, b. 1-T phase - $CrI_3$ phase and c. 1-T phase - $CrGeTe_3$ phase, where, in each layer, magnetic cations are sandwiched by ligands at two sides.

## III. INTERLAYER EXCHANGE PATHWAYS

For insulating type I-I bilayers, the interlayer exchange pathway for other electronic configurations, including $t_{2g}^1 e_g^0 - t_{2g}^1 e_g^0$, $t_{2g}^2 e_g^0 - t_{2g}^2 e_g^0$, and $t_{2g}^3 e_g^0 - t_{2g}^3 e_g^0 e_g^1$, are presented in Figure S4. Since we focus on insulators, the band gap between occupied and empty $d$ orbitals is assumed. And this band gap opening can be realized by either onsite Coulomb repulsion or crystal field distortion [1, 2]. While for other electronic states in type II-II and type I-II, the exchange pathway follows the similar analysis.

* binghai.yan@weizmann.ac.il

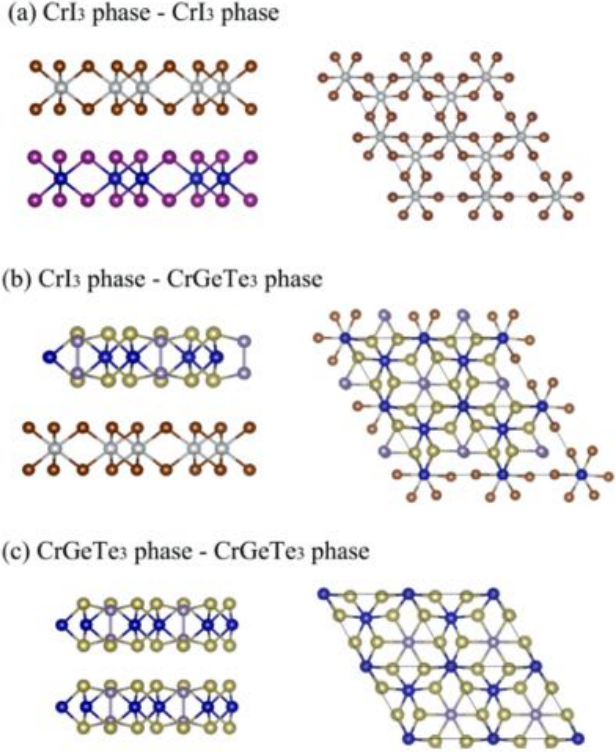

FIG. S3. $(0,0)$ stacking geometries for a. CrI$_3$ phase - CrI$_3$ phase, b. CrI$_3$ phase - CrGeTe$_3$ phase and c. CrGeTe$_3$ phase - CrGeTe$_3$ phase, where, in each layer, magnetic cations are sandwiched by ligands at two sides.

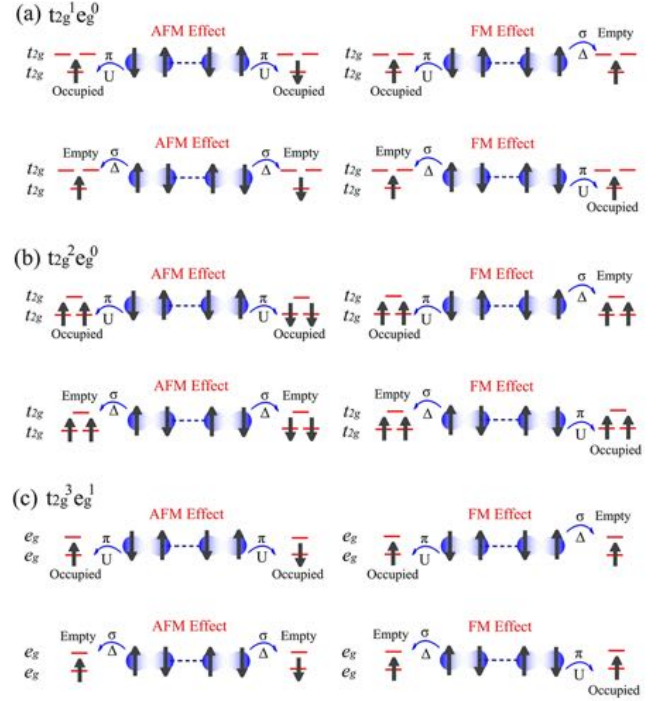

FIG. S4. Exchange pathways for a. $t_{2g}^1 e_g^0$-$t_{2g}^1 e_g^0$, b. $t_{2g}^2 e_g^0$-$t_{2g}^2 e_g^0$ and c. $t_{2g}^3 e_g^1$-$t_{2g}^3 e_g^1$, where the band gap between occupied and empty $d$ orbitals is assumed

## IV. DOPED INSULATING LAYERS

Electron and hole doped insulating bilayers, including type I-I CrI$_3$-CrI$_3$, type II-II NiCl$_2$-NiCl$_2$, and type I-II MnS$_2$-MnBr$_2$, are calculated and presented in Figure S5. It shows that, with the increasing carrier concentration, the AFM coupling for both type I-I CrI$_3$-CrI$_3$ and type II-II NiCl$_2$-NiCl$_2$ is gradually weakened, and the phase transition from AFM to FM can be realized. For type I-II MnS$_2$-MnBr$_2$, electron doping brings additional AFM effect due to the different spin components of conducting electrons in different layers. While for hole doping, free carriers with the same spin still enhance the FM coupling.

To understand the FM effect brought by the hole doping in type II-II, the band structure for type I-II MnS$_2$-NiBr$_2$ (shown in the main text) is presented in Figure S6. It indicates that, bands right below Fermi-level are filled by spin up electrons for both type I and type II layer. Therefore, when the Fermi level is shifted down, introduced holes in two layers belong to the same spin up channel, in contrary to the electron doping.

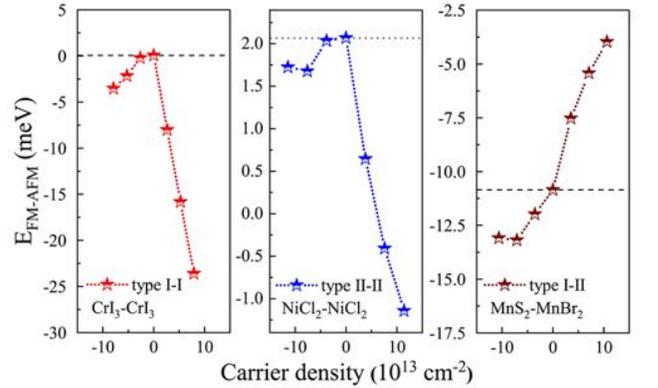

FIG. S5. The evolution of E$_{FM-AFM}$ against carrier density for type I-I CrI$_3$-CrI$_3$, type II-II NiCl$_2$-NiCl$_2$ and type I-II MnS$_2$-MnBr$_2$

## V. MODEL HAMILTONIAN FOR INTERLAYER MAGNETIC COUPLING

In the main text, we construct a phenomenological model Hamiltonian and will solve it analytically in the

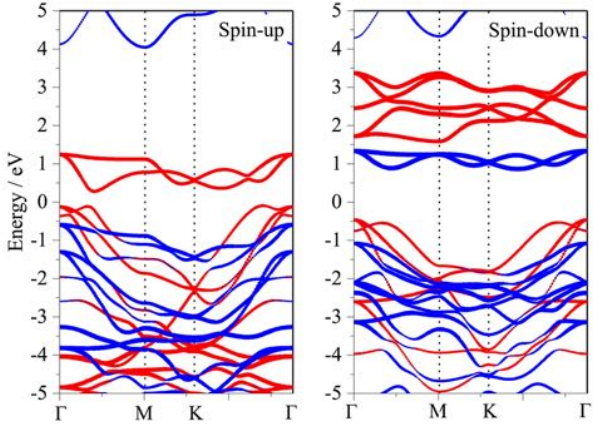

FIG. S6. Band structure of type I-II MnS$_2$-NiBr$_2$, where red and blue dots represent the total orbital weight of type I and type II layer respectively.

following.

$$H = \sum_{i,l,u,\sigma} \epsilon_{lu\sigma} c^\dagger_{ilu\sigma} c_{ilu\sigma} - t \sum_{ij,l,uv,\sigma} (c^\dagger_{ilu\sigma} c_{jlv\sigma} + H.C.)$$
$$+ U \sum_{i,l,u} n_{ilu\uparrow} n_{ilu\downarrow} - t_l \sum_{i,l,u,\sigma} (c^\dagger_{ilu\sigma} c_{i(1-l)u\sigma} + H.C.) \quad (1)$$

As stated in the main text, four subsequent terms correspond to onsite energy, kinetic energy of intralayer electron hopping ($t$), onsite Coulomb repulsion ($U$) and kinetic energy of interlayer electron hopping ($t_l$). Here $c^\dagger_{ilu\sigma}$ ($c_{ilu\sigma}$) is the electron creation (annihilation) operator for orbital $u$ at site $i$ and layer $l$ with spin $\sigma$. $l$ has the value 0 and 1, denoting the lower and upper layers. $\epsilon_{lu\sigma}$ is the onsite energy for orbital $u$ ($u = d, p$) with spin $\sigma$ at layer $l$. Since the interlayer exchange is along the vertical direction without periodicity, our discussions focus on the same site within a unit cell. Therefore, the intralayer site index $i$ is removed and the orbital basis involved in the interlayer hopping, as shown in Figure 2, can be generally expressed as $c^\dagger_i = (c^\dagger_{0d\uparrow}, c^\dagger_{0p\uparrow}, c^\dagger_{0d\downarrow}, c^\dagger_{0p\downarrow}, c^\dagger_{1d\uparrow}, c^\dagger_{1p\uparrow}, c^\dagger_{1d\downarrow}, c^\dagger_{1p\downarrow})$. Thus, the matrix form of Hamiltonian is derived:

$$H = \begin{pmatrix} H_{00} & O_{01} \\ O_{10} & H_{11} \end{pmatrix} \quad (2)$$

$$H_{ll} = \begin{pmatrix} E_{ld\uparrow} & t & 0 & 0 \\ t & E_{lp\uparrow} & 0 & 0 \\ 0 & 0 & E_{ld\downarrow} & t \\ 0 & 0 & t & E_{lp\downarrow} \end{pmatrix} \quad (3)$$

$$O_{01FM} = \begin{pmatrix} 0 & 0 & 0 & 0 \\ 0 & t_l & 0 & 0 \\ 0 & 0 & 0 & 0 \\ 0 & 0 & 0 & t_l \end{pmatrix} \quad (4)$$

$$O_{01AFM} = \begin{pmatrix} 0 & 0 & 0 & 0 \\ 0 & 0 & 0 & t_l \\ 0 & 0 & 0 & 0 \\ 0 & t_l & 0 & 0 \end{pmatrix} \quad (5)$$

where $O_{01FM}$ and $O_{01AFM}$ represent interlayer hopping matrix for FM and AFM order respectively. $E_{lu\sigma}$ ($u = d, p; l = 0, 1; \sigma = \uparrow, \downarrow$) incorporates the onsite energy $\epsilon_{lu\sigma}$ and Coulomb repulsion.The electron hopping among the nearest $d$ and $p$ orbitals is simplified as $t$ [3], which is much larger than $t_l$. To evaluate the weak interlayer electron hopping, each pair of intralayer $|d_{l\sigma}\rangle$ and $|p_{l\sigma}\rangle$ orbitals are transformed into hybridized orbitals as $\alpha_{l\sigma} |d_{l\sigma}\rangle + \beta_{l\sigma} |p_{l\sigma}\rangle$ and $\alpha_{l\sigma} |p_{l\sigma}\rangle - \beta_{l\sigma} |d_{l\sigma}\rangle$ ($\alpha_{l\sigma} = \frac{1}{\sqrt{1 + \frac{t_\sigma^2}{(E_{pl\sigma} - E_{dl\sigma})^2}}}, \alpha_{l\sigma}^2 + \beta_{l\sigma}^2 = 1, \alpha_{l\sigma} > \beta_{l\sigma}$), and the corresponding Hamiltonian is transformed into:

$$H = \begin{pmatrix} H_{00} & O_{01} \\ O_{10} & H_{11} \end{pmatrix} \quad (6)$$

$$H_{ll} = \begin{pmatrix} E'_{ld\uparrow} & 0 & 0 & 0 \\ 0 & H'_{lp\uparrow} & 0 & 0 \\ 0 & 0 & E'_{ld\downarrow} & 0 \\ 0 & 0 & 0 & E'_{lp\downarrow} \end{pmatrix} \quad (7)$$

$$O_{01FM} = \begin{pmatrix} \beta_{0\uparrow}\beta_{1\uparrow}t_l & \beta_{0\uparrow}\alpha_{1\uparrow}t_l & 0 & 0 \\ \alpha_{0\uparrow}\beta_{1\uparrow}t_l & \alpha_{0\uparrow}\alpha_{1\uparrow}t_l & 0 & 0 \\ 0 & 0 & \beta_{0\downarrow}\beta_{1\downarrow}t_l & \beta_{0\downarrow}\alpha_{1\downarrow}t_l \\ 0 & 0 & \alpha_{0\downarrow}\beta_{1\downarrow}t_l & \alpha_{0\downarrow}\alpha_{1\downarrow}t_l \end{pmatrix} \quad (8)$$

$$O_{01AFM} = \begin{pmatrix} 0 & 0 & \beta_{0\uparrow}\beta_{1\downarrow}t_l & \beta_{0\uparrow}\alpha_{1\downarrow}t_l \\ 0 & 0 & \alpha_{0\uparrow}\beta_{1\downarrow}t_l & \alpha_{0\uparrow}\alpha_{1\downarrow}t_l \\ \beta_{0\downarrow}\beta_{1\uparrow}t_l & \beta_{0\downarrow}\alpha_{1\uparrow}t_l & 0 & 0 \\ \alpha_{0\downarrow}\beta_{1\uparrow}t_l & \alpha_{0\downarrow}\alpha_{1\uparrow}t_l & 0 & 0 \end{pmatrix} \quad (9)$$

Where $E'_{ld\sigma} = E_{ld\sigma} + \frac{t^2}{E_{ld\sigma} - E_{lp\sigma}}, E'_{lp\sigma} = E_{lp\sigma} - \frac{t^2}{E_{ld\sigma} - E_{lp\sigma}}$ (We still use $E_{ld\sigma}$ and $E_{lp\sigma}$ to denote the hybridized onsite energies $E'_{ld\sigma}$ and $E'_{ld\sigma}$, for the simplicity of notation). It is worth to note that $\alpha_{l\sigma}$ represents the degree of intralayer $d$-$p$ hybridization, where the less the value of $\alpha_{l\sigma}$ reflects the stronger intralayer hybridization. Furthermore, $2(\alpha_\downarrow - \alpha_\uparrow)$ can be approximated as the induced magnetic moment on ligands. Since $p$ orbitals hybridize with empty $d$ orbitals, they are partially occupied by $\alpha_{l\sigma}^2$ percent. Therefore, the induced magnetic moment $m$ is expressed as $(\alpha_{l\uparrow}^2 - \alpha_{l\downarrow}^2)\mu_B \approx 2(\alpha_{l\uparrow} - \alpha_{l\downarrow})\mu_B$, and it will be used as a parameter in the phase diagram. On the other hand, for the interlayer parameters, the interlayer hopping integral further face the geometry of orbital orientation. For instance, empty $e_g$ hybridizes with $p$ orbitals via $\sigma$ bond, and we denote such $p$ orbitals as $p_\sigma$. While $t_{2g}$ hybridizes with $p_\pi$ orbitals through $\pi$ bond. As a result, different intralayer $p$ orbitals with varied geometries interact with each other across the van der Waals

gap. Thus, $t_l$ can be classified into three types of interlayer hopping process: $t_{p_\sigma - p_\sigma}$, $t_{p_\sigma - p_\pi}$, and $t_{p_\sigma - p_\sigma}$, and their expressions are detailed in the following Section VI.

For the mathematical convenience, we further reorganize the orbital basis, where former and later four orbitals belong to spin up and spin down, respectively. For FM order, the Hamiltonian is expressed as:

$$H_{FM} = \begin{pmatrix} H_{FM\uparrow} & O \\ O & H_{FM\downarrow} \end{pmatrix} \tag{10}$$

$$H_{FM\sigma} = \begin{pmatrix} E_{0d\sigma} & 0 & \beta_{0\sigma}\beta_{1\sigma}t_l & \beta_{0\sigma}\alpha_{1\sigma}t_l \\ 0 & E_{0p\sigma} & \alpha_{0\sigma}\beta_{1\sigma}t_l & \alpha_{0\sigma}\alpha_{1\sigma}t_l \\ \beta_{0\sigma}\beta_{1\sigma}t_l & \alpha_{0\sigma}\beta_{1\sigma}t_l & E_{1d\sigma} & 0 \\ \beta_{0\sigma}\alpha_{1\sigma}t_l & \alpha_{0\sigma}\alpha_{1\sigma}t_l & 0 & E_{1p\sigma} \end{pmatrix} \tag{11}$$

Similarly, for AFM order, the Hamiltonian is changed into:

$$H_{AFM} = \begin{pmatrix} H_{AFM\uparrow} & O \\ O & H_{AFM\downarrow} \end{pmatrix} \tag{12}$$

$$H_{AFM\sigma} = \begin{pmatrix} E_{0d\sigma} & 0 & \beta_{0\sigma}\beta_{1\sigma'}t_l & \beta_{0\sigma}\alpha_{1\sigma'}t_l \\ 0 & E_{0p\sigma} & \alpha_{0\sigma}\beta_{1\sigma'}t_l & \alpha_{0\sigma}\alpha_{1\sigma'}t_l \\ \beta_{0\sigma}\beta_{1\sigma'}t_l & \alpha_{0\sigma}\beta_{1\sigma'}t_l & E_{1d\sigma'} & 0 \\ \beta_{0\sigma}\alpha_{1\sigma'}t_l & \alpha_{0\sigma}\alpha_{1\sigma'}t_l & 0 & E_{1p\sigma'} \end{pmatrix} \tag{13}$$

### A. type I-I

For type I-I bilayer, the energy difference between interlayer FM and AFM order can be derived based on the second order perturbation,:

$$
\begin{aligned}
E_{FM-AFM} = & t_{AFM\uparrow}^2 (\alpha_{0\uparrow}^2 \frac{1-\alpha_{1\downarrow}^2}{\Delta_{1\downarrow}} + \alpha_{1\uparrow}^2 \frac{1-\alpha_{0\downarrow}^2}{\Delta_{0\downarrow}}) \\
& + t_{AFM\downarrow}^2 (\alpha_{0\downarrow}^2 \frac{1-\alpha_{1\uparrow}^2}{\Delta_{1\uparrow}} + \alpha_{1\downarrow}^2 \frac{1-\alpha_{0\uparrow}^2}{\Delta_{0\uparrow}}) \\
& - t_{FM\uparrow}^2 (\alpha_{0\uparrow}^2 \frac{1-\alpha_{1\uparrow}^2}{\Delta_{1\uparrow}} + \alpha_{1\uparrow}^2 \frac{1-\alpha_{0\uparrow}^2}{\Delta_{0\uparrow}}) \\
& - t_{FM\downarrow}^2 (\alpha_{0\downarrow}^2 \frac{1-\alpha_{1\downarrow}^2}{\Delta_{1\downarrow}} + \alpha_{1\downarrow}^2 \frac{1-\alpha_{0\downarrow}^2}{\Delta_{0\downarrow}})
\end{aligned} \tag{14}
$$

Where the charge transfer gap $\Delta_{l\sigma}$ is defined as $E_{ld\sigma} - E_{lp\sigma}$. For the typical $t_{2g}^3 e_g^0$-$t_{2g}^3 e_g^0$ states, $p_\uparrow$ hybridizes with $e_g$ (denoted $p_\sigma$ type) and $p_\downarrow$ hybridizes with $t_{2g}$ (denoted $p_\pi$ type). Thus, the interlayer interaction in $H_{FM\uparrow}$, $H_{FM\downarrow}$, $H_{AFM\uparrow}$ and $H_{AFM\downarrow}$ block belongs to $p_\sigma$-$p_\sigma$, $p_\pi$-$p_\pi$, $p_\pi$-$p_\sigma$, $p_\sigma$-$p_\pi$ type, with $t_l$ expressed as $t_{p_\sigma - p_\sigma}, t_{p_\pi - p_\pi}, t_{p_\pi - p_\sigma}, t_{p_\sigma - p_\pi}$, respectively. As a result,

$E_{FM-AFM}$ is modified as:

$$
\begin{aligned}
E_{FM-AFM} = & t_{p_\sigma - p_\sigma}^2 (\alpha_{0\uparrow}^2 \frac{1-\alpha_{1\downarrow}^2}{\Delta_{1\downarrow}} + \alpha_{1\uparrow}^2 \frac{1-\alpha_{0\downarrow}^2}{\Delta_{0\downarrow}}) \\
& + t_{p_\pi - p_\sigma}^2 (\alpha_{0\downarrow}^2 \frac{1-\alpha_{1\uparrow}^2}{\Delta_{1\uparrow}} + \alpha_{1\downarrow}^2 \frac{1-\alpha_{0\uparrow}^2}{\Delta_{0\uparrow}}) \\
& - t_{p_\sigma - p_\sigma}^2 (\alpha_{0\uparrow}^2 \frac{1-\alpha_{1\uparrow}^2}{\Delta_{1\uparrow}} + \alpha_{1\uparrow}^2 \frac{1-\alpha_{0\uparrow}^2}{\Delta_{0\uparrow}}) \\
& - t_{p_\pi - p_\pi}^2 (\alpha_{0\downarrow}^2 \frac{1-\alpha_{1\downarrow}^2}{\Delta_{1\downarrow}} + \alpha_{1\downarrow}^2 \frac{1-\alpha_{0\downarrow}^2}{\Delta_{0\downarrow}})
\end{aligned} \tag{15}
$$

Therefore, $E_{FM-AFM}$ exhibits the competing effect, and is easily to be affected by the interlayer hopping integral and thus the stacking order. For the simplified homobilayer phase (where layer index $l$ can be removed), we further plotted the contour map of $E_{FM-AFM}$ against intralayer parameters $m = 2(\alpha_\downarrow - \alpha_\uparrow)$ and charge transfer gap $\Delta_\sigma$, as shown in the main text (Detailed approximations are presented in the following Section VI).

### B. type II-II

For type II-II bilayer, the interlayer electron hopping involve six orbitals (four $p$ orbitals in upper and lower layers with different spins and two spin down $d$ orbitals in upper and lower layers), since there are no empty spin up $d$ orbitals available for type II layer. The orbital basis and Hamiltonian can be expressed as:

$$
\begin{aligned}
& |p_{0\uparrow}\rangle, |p_{1\uparrow}\rangle \\
\alpha_{0\downarrow} |d_{0\downarrow}\rangle + \beta_{0\downarrow} |p_{0\downarrow}\rangle, & \alpha_{0\downarrow} |p_{0\downarrow}\rangle - \beta_{0\downarrow} |d_{0\downarrow}\rangle \\
\alpha_{1\downarrow} |d_{1\downarrow}\rangle + \beta_{1\downarrow} |p_{1\downarrow}\rangle, & \alpha_{1\downarrow} |p_{1\downarrow}\rangle - \beta_{1\downarrow} |d_{1\downarrow}\rangle
\end{aligned} \tag{16}
$$

$$H_{FM} = \begin{pmatrix} H_{FM\uparrow} & O \\ O & H_{FM\downarrow} \end{pmatrix} \tag{17}$$

$$H_{FM\uparrow} = \begin{pmatrix} E_{0p\uparrow} & t_l \\ t_l & E_{1p\uparrow} \end{pmatrix} \tag{18}$$

$$H_{FM\downarrow} = \begin{pmatrix} E_{0d\downarrow} & 0 & \beta_{0\downarrow}\beta_{1\downarrow}t_l & \beta_{0\downarrow}\alpha_{1\downarrow}t_l \\ 0 & E_{0p\downarrow} & \alpha_{0\downarrow}\beta_{1\downarrow}t_l & \alpha_{0\downarrow}\alpha_{1\downarrow}t_l \\ \beta_{0\downarrow}\beta_{1\downarrow}t_l & \alpha_{0\downarrow}\beta_{1\downarrow}t_l & E_{1d\downarrow} & 0 \\ \beta_{0\downarrow}\alpha_{1\downarrow}t_l & \alpha_{0\downarrow}\alpha_{1\downarrow}t_l & 0 & E_{1p\downarrow} \end{pmatrix} \tag{19}$$

$$H_{AFM} = \begin{pmatrix} H_{AFM\uparrow} & O \\ O & H_{AFM\downarrow} \end{pmatrix} \tag{20}$$

$$H_{AFM\uparrow} = \begin{pmatrix} E_{0p\uparrow} & \beta_{1\downarrow}t_l & \alpha_{1\downarrow}t_l \\ \beta_{1\downarrow}t_l & E_{1d\downarrow} & 0 \\ \alpha_{1\downarrow}t_l & 0 & E_{1p\downarrow} \end{pmatrix} \tag{21}$$

$$H_{AFM\downarrow} = \begin{pmatrix} E_{0d\downarrow} & 0 & \beta_{0\downarrow}t_l \\ 0 & E_{0p\downarrow} & \alpha_{0\downarrow}t_l \\ \beta_{0\downarrow}t_l & \alpha_{0\downarrow}t_l & E_{1p\uparrow} \end{pmatrix} \quad (22)$$

Since $p_\uparrow$ in type II does not hybridize with spin up $d$ orbitals with no selected geometry (denoted as $p$ type), while $p_\downarrow$ interacts with $t_{2g}$ (denoted $p_\pi$ type), the interlayer interaction in $H_{FM\uparrow}$, $H_{FM\downarrow}$, $H_{AFM\uparrow}$ and $H_{AFM\downarrow}$ block belongs to $p$-$p$, $p_\pi$-$p_\pi$, $p_\pi$-$p_\pi$, $p_\pi$-$p$ type, with $t_l$ expressed as $t_{p-p}$, $t_{p_\pi-p_\pi}$, $t_{p-p_\pi}$, $t_{p_\pi-p}$. However, we later show that, $t_{p-p_{\pi(\sigma)}}$ can be approximated as $t_{p_\pi-p_\pi}$ (see in the following Section VI). Therefore, based on the second order perturbation, the energy difference between FM and AFM order is derived:

$$E_{FM-AFM} = t_{p_\pi-p_\pi}^2 \left( \frac{1-\alpha_{1\downarrow}^2}{\Delta_{1\downarrow}} (1-\alpha_{0\downarrow}^2) \right. \\ \left. + \frac{1-\alpha_{0\downarrow}^2}{\Delta_{0\downarrow}} (1-\alpha_{1\downarrow}^2) \right) \quad (23)$$

Due to $\alpha_{0\downarrow}, \alpha_{0\uparrow} \leq 1$, AFM order is always preferred, as predicted. For the simplified homo-bilayer phase, Figure 4(b) further plots the contour map of $E_{FM-AFM}$, with approximations detailed in the following section.

## C. I-II

For type I-II bilayer with the lower layer as type I and the topper layer as type II, the interlayer electron hopping involve seven orbitals, since there are no empty spin up $d$ orbitals in type II layer. The orbital basis and Hamiltonian are expressed as:

$$\alpha_{0\uparrow}|d_{0\uparrow}\rangle + \beta_{0\uparrow}|p_{0\uparrow}\rangle, \alpha_{0\uparrow}|p_{0\uparrow}\rangle - \beta_{0\uparrow}|d_{0\uparrow}\rangle \\ |p_{1\uparrow}\rangle \\ \alpha_{0\downarrow}|d_{0\downarrow}\rangle + \beta_{0\downarrow}|p_{0\downarrow}\rangle, \alpha_{0\downarrow}|p_{0\downarrow}\rangle - \beta_{0\downarrow}|d_{0\downarrow}\rangle \\ \alpha_{1\downarrow}|d_{1\downarrow}\rangle + \beta_{1\downarrow}|p_{1\downarrow}\rangle, \alpha_{1\downarrow}|p_{1\downarrow}\rangle - \beta_{1\downarrow}|d_{1\downarrow}\rangle \quad (24)$$

$$H_{FM} = \begin{pmatrix} H_{FM\uparrow} & O \\ O & H_{FM\downarrow} \end{pmatrix} \quad (25)$$

$$H_{FM\uparrow} = \begin{pmatrix} E_{0d\uparrow} & 0 & \beta_{0\uparrow}t_l \\ 0 & E_{0p\uparrow} & \alpha_{0\uparrow}t_l \\ \beta_{0\uparrow}t_l & \alpha_{0\uparrow}t_l & E_{1p\uparrow} \end{pmatrix} \quad (26)$$

$$H_{FM\downarrow} = \begin{pmatrix} E_{0d\downarrow} & 0 & \beta_{0\downarrow}\beta_{1\downarrow}t_l & \beta_{0\downarrow}\alpha_{1\downarrow}t_l \\ 0 & E_{0p\downarrow} & \alpha_{0\downarrow}\beta_{1\downarrow}t_l & \alpha_{0\downarrow}\alpha_{1\downarrow}t_l \\ \beta_{0\downarrow}\beta_{1\downarrow}t_l & \alpha_{0\downarrow}\beta_{1\downarrow}t_l & E_{1d\downarrow} & 0 \\ \beta_{0\downarrow}\alpha_{1\downarrow}t_l & \alpha_{0\downarrow}\alpha_{1\downarrow}t_l & 0 & E_{1p\downarrow} \end{pmatrix} \quad (27)$$

$$H_{AFM} = \begin{pmatrix} H_{AFM\uparrow} & O \\ O & H_{AFM\downarrow} \end{pmatrix} \quad (28)$$

$$H_{AFM\uparrow} = \begin{pmatrix} E_{0d\uparrow} & 0 & \beta_{0\uparrow}\beta_{1\downarrow}t_l & \beta_{0\uparrow}\alpha_{1\downarrow}t_l \\ 0 & E_{0p\uparrow} & \alpha_{0\uparrow}\beta_{1\downarrow}t_l & \alpha_{0\uparrow}\alpha_{1\downarrow}t_l \\ \beta_{0\uparrow}\beta_{1\downarrow}t_l & \alpha_{0\uparrow}\beta_{1\downarrow}t_l & E_{1d\downarrow} & 0 \\ \beta_{0\uparrow}\alpha_{1\downarrow}t_l & \alpha_{0\uparrow}\alpha_{1\downarrow}t_l & 0 & E_{1p\downarrow} \end{pmatrix} \quad (29)$$

$$H_{AFM\downarrow} = \begin{pmatrix} E_{0d\downarrow} & 0 & \beta_{0\downarrow}t_l \\ 0 & E_{0p\downarrow} & \alpha_{0\downarrow}t_l \\ \beta_{0\downarrow}t_l & \alpha_{0\downarrow}t_l & E_{1p\uparrow} \end{pmatrix} \quad (30)$$

In type I layer, $p_\uparrow$ that hybridizes with $e_g$ belongs to $p_\sigma$ type; $p_\downarrow$ that hybridizes with $t_{2g}$ belongs to $p_\pi$. In type II layer, $p_\uparrow$ without hybridization belongs to $p$ type; $p_\downarrow$ that interacts with $t_{2g}$ belongs to $p_\pi$. Therefore, the interlayer interaction in $H_{FM\uparrow}$, $H_{FM\downarrow}$, $H_{AFM\uparrow}$ and $H_{AFM\downarrow}$ blocks is classified into $p_\sigma$-$p$, $p_\pi$-$p_\pi$, $p_\sigma$-$p_\pi$, $p_\pi$-$p$ type, with $t_l$ expressed as $t_{p_\sigma-p}$, $t_{p_\pi-p_\pi}$, $t_{p_\sigma-p_\pi}$, $t_{p_\pi-p}$, respectively. Here, again, $t_{p-p_{\pi(\sigma)}}$ is approximated as $t_{p_\pi-p_\pi}$ (Section VI). Therefore, the energy difference between FM and AFM order can be derived as:

$$E_{FM-AFM} = (t_{p_\sigma-p_\pi}^2 \alpha_{1\downarrow}^2 - t_{p_\pi-p_\pi}^2) \frac{1-\alpha_{0\uparrow}^2}{\Delta_{0\uparrow}} \\ + (t_{p_\sigma-p_\pi}^2 \alpha_{0\uparrow}^2 - t_{p_\pi-p_\pi}^2 \alpha_{0\downarrow}^2) \frac{1-\alpha_{1\downarrow}^2}{\Delta_{1\downarrow}} \\ - (t_{p_\pi-p_\pi}^2 \alpha_{1\downarrow}^2 - t_{p_\pi-p_\pi}^2) \frac{1-\alpha_{0\downarrow}^2}{\Delta_{0\downarrow}} \quad (31)$$

Due to $\alpha_{l\sigma} \leq 1$, $\alpha_{0\uparrow} < \alpha_{0\downarrow}$ and $t_{p_\sigma-p_\pi} < t_{p_\pi-p_\pi}$, the first and second term favors FM order while the third term is AFM. Thus, competition exists, as expected. Comparing the first and third terms, it shows that FM effect prevails: Both $\Delta_{0\uparrow}$ and $\alpha_{0\uparrow}^2$ are smaller than $\Delta_{0\downarrow}$ and $\alpha_{0\downarrow}^2$ due to the exchange splitting in type I layer, and thus the larger value of $\frac{1-\alpha_{0\uparrow}^2}{\Delta_{0\uparrow}}$ compared to $\frac{1-\alpha_{0\downarrow}^2}{\Delta_{0\downarrow}}$ enhances the FM contribution. In combination with the additional FM effect from the second term, interlayer FM order dominates. Corresponding contour map is plotted in Figure 4(c) and 4(d) in the main text, with approximations shown in the following section.

## D. Biaxial strain modulation

Based on the contour map, the modulation of the interlayer coupling strength is explored, and here we focus on the biaxial strain. For type I-I $MnS_2$-$MnS_2$, results in Figure S7(a) show that, a slight compression strain can reduce the spin polarization $|m|$ on ligands and thus lead to the weakened exchange splitting (along with the negligible variances of $\Delta_{l\sigma}$). And the combined effect is the enhanced FM coupling according to Figure 4(a). For type II-II $MnCl_2$-$MnCl_2$ in Figure S7(b), when the compression biaxial strain is applied, the ligand polarization $|m|$ is enhanced and further strengthens the AFM coupling,

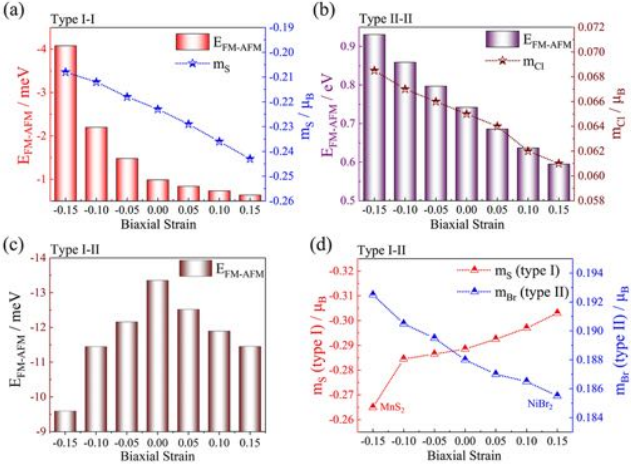

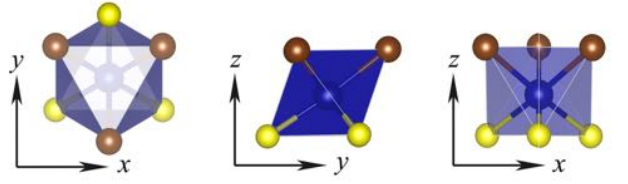

FIG. S8. The crystal geometry of magnetic monolayers

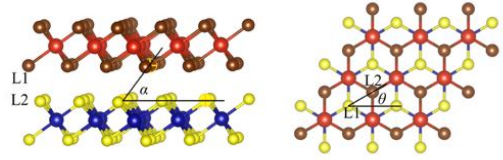

FIG. S9. The interlayer geometry, where L1 and L2 denote ligands in different layers.

FIG. S7. The evolution of $E_{FM-AFM}$ and magnetic moment on ligands against biaxial strain for a. $MnS_2$-$MnS_2$ (type I-I), b. $MnCl_2$-$MnCl_2$ (type II-II) bilayer, and c, d. $MnS_2$-$NiBr_2$ (type I-II) bilayer.

being consistent with Figure 4(b). While for type I-II $MnS_2$-$NiBr_2$ with multiple parameters, results in Figure S7(c) and S7(d) show that compression strain increases the ligand polarization $|m|$ of type II layer, and favors FM coupling, in contrary to the enhanced AFM effect brought by the reduced $|m|$ in type I. Therefore, competition exists, but the later contribution is stronger and can thus decrease the FM coupling in the compression region. While for extension strain, the AFM effect from the decreasing $|m|$ in type II dominates and also weakens the intrinsic FM coupling.

## VI. INTERLAYER HOPPING INTEGRAL FOR CONTOUR MAPS

To plot the contour map, the interlayer hopping integral $t_l$ is derived based on the two center Slater-Koster approximation [4]. The crystal geometry of magnetic layers are presented in Figure S8. Under this coordination, $p_\pi$ type ($p$ orbital that hybridizes with $t_{2g}$) and $p_\sigma$ type ($p$ orbital that hybridizes with $e_g$) orbitals are expressed as:

$$|p_{\pi 1}\rangle = (\sqrt{\frac{1}{3}}, 0, \sqrt{\frac{2}{3}})$$
$$|p_{\pi 2}\rangle = (-\frac{1}{2}\sqrt{\frac{1}{3}}, -\frac{1}{2}, \sqrt{\frac{2}{3}})$$
$$|p_{\pi 3}\rangle = (-\frac{1}{2}\sqrt{\frac{1}{3}}, \frac{1}{2}, \sqrt{\frac{2}{3}})$$
$$|p_{\sigma 1}\rangle = (1, 0, 0)$$
$$|p_{\sigma 2}\rangle = (0, -1, 0)$$

(32)

We choose $|p_{\pi 1}\rangle$ and $|p_{\sigma 1}\rangle$ as a typical representative,

and express $t_{p_\sigma - p_\pi}$, $t_{p_\pi - p_\pi}$, $t_{p_\sigma - p_\pi}$ as:

$$t_{p_\pi - p_\pi} = \frac{1}{3}t_{p_x - p_x} + \frac{2}{3}t_{p_z - p_z} + \frac{2\sqrt{2}}{3}t_{p_x - p_z}$$
$$t_{p_\pi - p_\sigma} = \sqrt{\frac{1}{3}}t_{p_x - p_x} + \sqrt{\frac{2}{3}}t_{p_x - p_z}$$
$$t_{p_\sigma - p_\sigma} = t_{p_x - p_x}$$

(33)

With the interlayer geometry shown in Figure S9 and two-center Slater-Koster approximation, $t_{p_x - p_z}$, $t_{p_x - p_x}$ and $t_{p_z - p_z}$ can be expressed as:

$$t_{p_x - p_x} = U_{pp\pi}sin^2(\theta) - U_{pp\sigma}cos^2(\theta)cos^2(\alpha)$$
$$+ U_{pp\pi}cos^2(\theta)sin^2(\alpha)$$
$$t_{p_z - p_z} = U_{pp\pi}cos^2(\alpha) - U_{pp\sigma}sin^2(\alpha)$$
$$t_{p_x - p_z} = -cos(\theta)sin(\alpha)cos(\alpha)(U_{pp\pi} + U_{pp\sigma})$$

(34)

$U_{pp\sigma}$ is approximated as $3U_{pp\pi}$ [5]. After averaging $\theta$ and adopting $\alpha$ as $\frac{\pi}{3}$, we derive the following approximation:

$$t_{p_\pi - p_\pi} = 2U_{pp\pi}$$
$$t_{p_\pi - p_\sigma} = U_{pp\pi}$$
$$t_{p_\sigma - p_\sigma} = 0.5U_{pp\pi}$$

(35)

While for the above $t_{p-p_{\pi(\sigma)}}$ type, we expect $p$ type orbitals without selected geometry try to maximize the interlayer interaction with $p_{\pi(\sigma)}$ states. Therefore, its strength is approximated as the strong one $t_{p_\pi - p_\pi}$. We also tested the weaker value of $t_{p-p_{\pi(\sigma)}}$ as $1.5 U_{\pi\pi}$, and plotted the contour map for type I-II bilayer. As shown in Figure S10(a) and S10(b), they exhibit the similar feature as that in Figure 4(c) and 4(d) in the main text.

With the above parameters for $t_l$, $E_{FM-AFM}$ of the

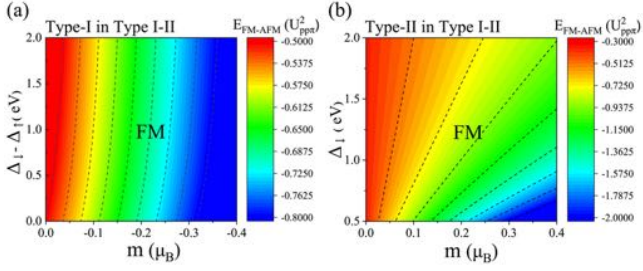

FIG. S10. The contour map of $E_{FM-AFM}$ against magnetic moment on ligands and charge transfer gap for type I-II bilayer, where $t_{p-p_\pi(\sigma)}$ is approximated as 1.5 $U_{\pi\pi}$.

homo-bilayer phase of type I-I becomes:

$$
\begin{aligned}
E_{FM-AFM} = U_{pp\pi}^2 (\frac{1-\alpha_\uparrow^2}{\Delta_\uparrow}(2\alpha_\downarrow^2 - 0.5\alpha_\uparrow^2) \\
+ \frac{1-\alpha_\downarrow^2}{\Delta_\downarrow}(2\alpha_\uparrow^2 - 8\alpha_\downarrow^2))
\end{aligned}
\tag{36}
$$

With the additional approximation as $\alpha_\downarrow = 0.95$ and $\Delta_\downarrow = 3eV$, the contour map against $m = 2(\alpha_\downarrow - \alpha_\uparrow)$ and $\Delta_\downarrow - \Delta_\uparrow$ can be plotted (Figure 4(a)). For the homo-bilayer phase of type II-II layer, $E_{FM-AFM}$ can be derived as:

$$
\begin{aligned}
E_{FM-AFM} = 4U_{pp\pi}^2 (\frac{1-\alpha_{1\downarrow}^2}{\Delta_{1\downarrow}}(1-\alpha_{0\downarrow}^2) \\
+ \frac{1-\alpha_{0\downarrow}^2}{\Delta_{0\downarrow}}(1-\alpha_{1\downarrow}^2))
\end{aligned}
\tag{37}
$$

Finally, for the type I-II bilayer, $E_{FM-AFM}$ is expressed as:

$$
\begin{aligned}
E_{FM-AFM} = U_{pp\pi}^2 (\alpha_{1\downarrow}^2 - 4)\frac{1-\alpha_{0\uparrow}^2}{\Delta_{0\uparrow}} \\
+ U_{pp\pi}^2 (\alpha_{0\uparrow}^2 - 4\alpha_{0\downarrow}^2)\frac{1-\alpha_{0\downarrow}^2}{\Delta_{1\downarrow}} \\
- 4U_{pp\pi}^2 (\alpha_{1\downarrow}^2 - 1)\frac{1-\alpha_{0\downarrow}^2}{\Delta_{0\downarrow}}
\end{aligned}
\tag{38}
$$

For parameters in type I layer, we additionally approximate $\alpha_{0\downarrow} = 0.95, \alpha_{1\downarrow} = 0.9, \Delta_{0\uparrow} = \Delta_{1\downarrow} = 1.5eV$, to plot the contour map in Figure 4(c). While for parameters in type II layer, approximation is adopted as $\alpha_{0\downarrow} = 0.80, \alpha_{0\downarrow} = 0.95, \Delta_{0\uparrow} = 1.5eV, \Delta_{0\downarrow} = 3.0eV$, to plot the contour map in Figure 4(d).

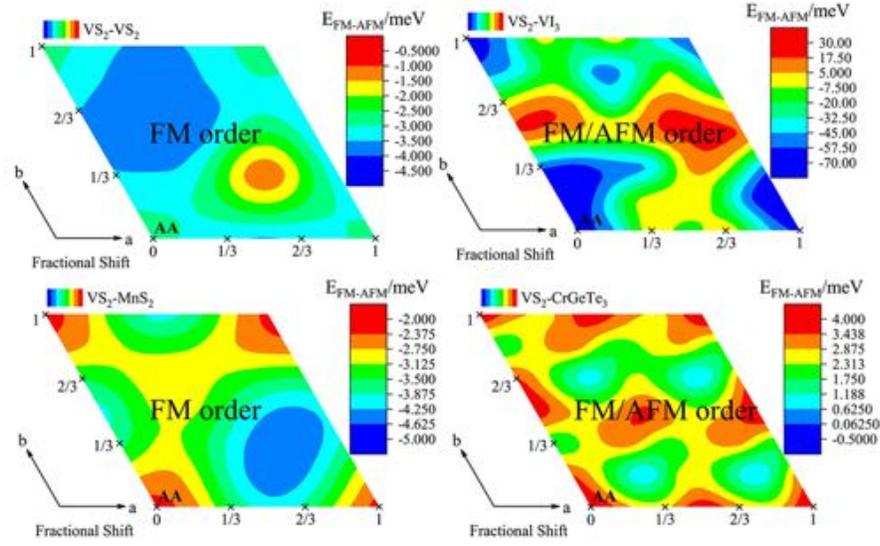

FIG. S11. The contour map for type I-I VS$_2$-VS$_2$, VS$_2$-VI$_3$, VS$_2$-MnS$_2$ and VS$_2$-CrGeTe$_3$ bilayers.

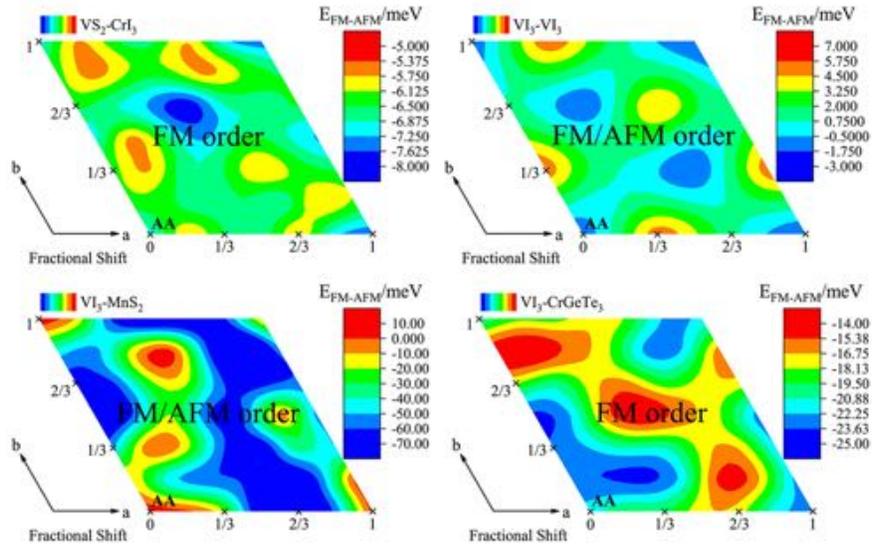

FIG. S12. The contour map for type I-I VS$_2$-CrI$_3$, VI$_3$-VI$_3$, VI$_3$-MnS$_2$ and VI$_3$-CrGeTe$_3$ bilayers.

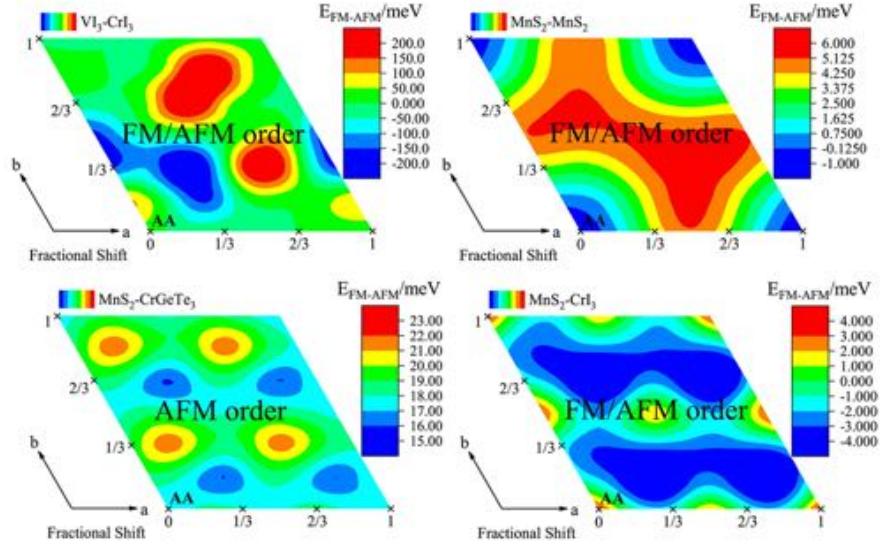

FIG. S13. The contour map for type I-I VI$_3$-CrI$_3$, MnS$_2$-MnS$_2$, MnS$_2$-CrGeTe$_3$ and MnS$_2$-CrI$_3$ bilayers.

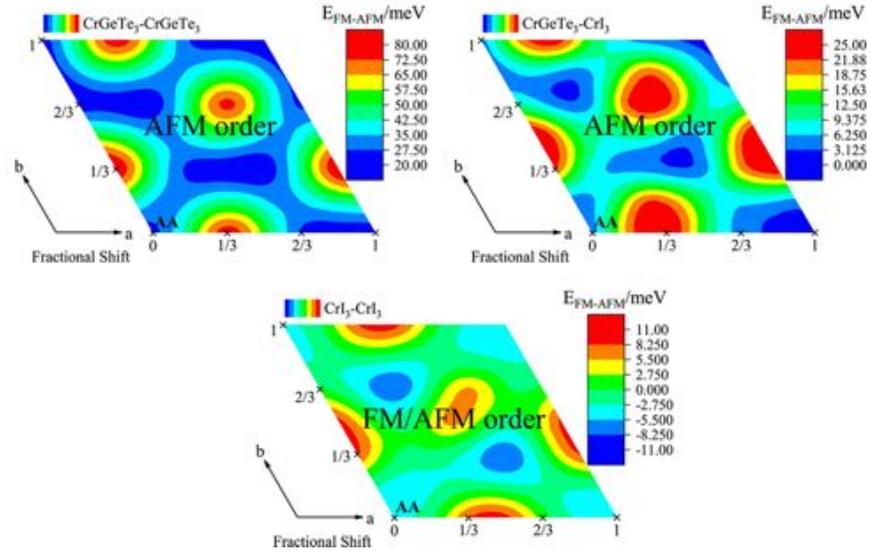

FIG. S14. The contour map for type I-I CrGeTe$_3$-CrGeTe$_3$, CrGeTe$_3$-CrI$_3$ and CrI$_3$-CrI$_3$ bilayers.

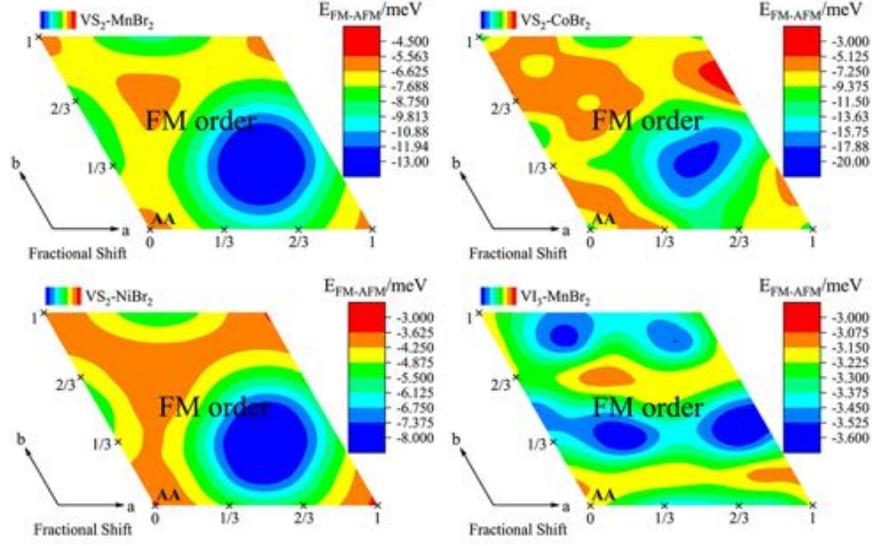

FIG. S15. The contour map for type I-II VS$_2$-MnBr$_2$, VS$_2$-CoBr$_2$, VS$_2$-NiBr$_2$ and VI$_3$-MnBr$_2$ bilayers.

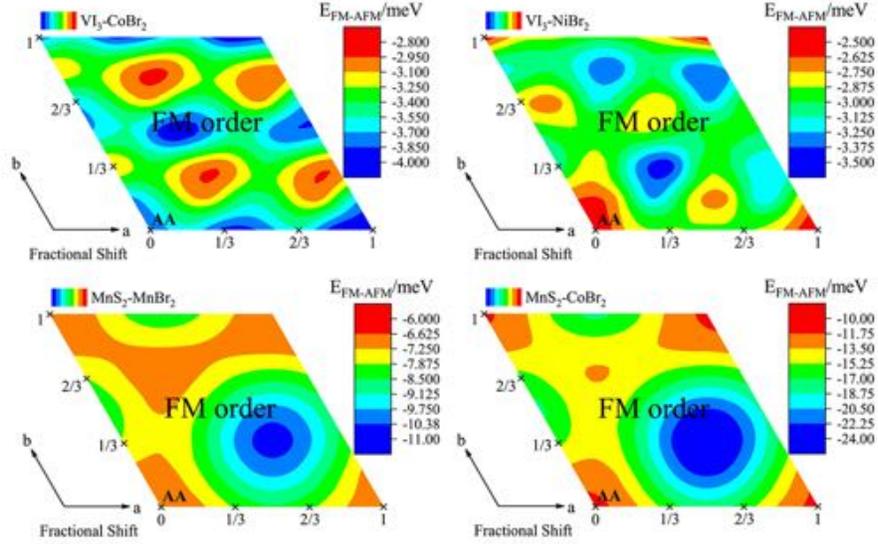

FIG. S16. The contour map for type I-II VI$_3$-CoBr$_2$, VI$_3$-NiBr$_2$, MnS$_2$-MnBr$_2$ and MnS$_2$-CoBr$_2$ bilayers.

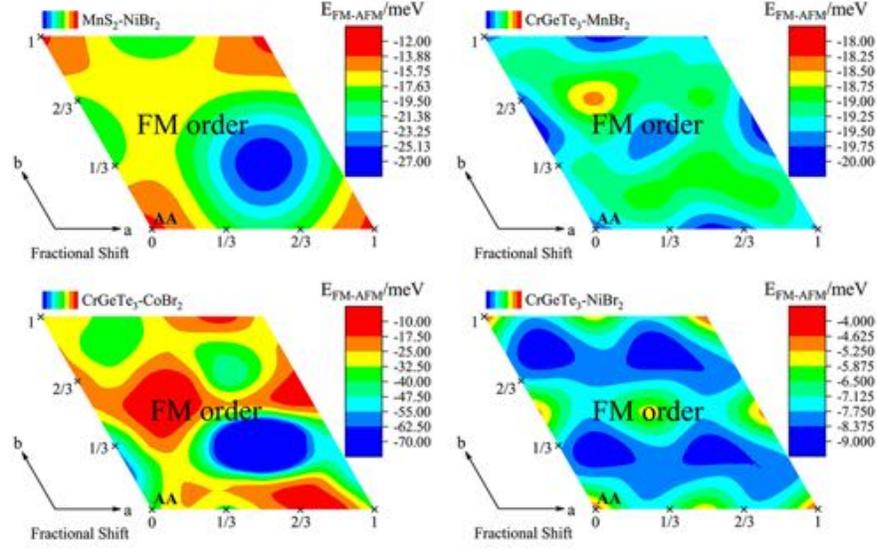

FIG. S17. The contour map for type I-II $MnS_2$-$NiBr_2$, $CrGeTe_3$-$MnBr_2$, $CrGeTe_3$-$CoBr_2$ and $CrGeTe_3$-$NiBr_2$ bilayers.

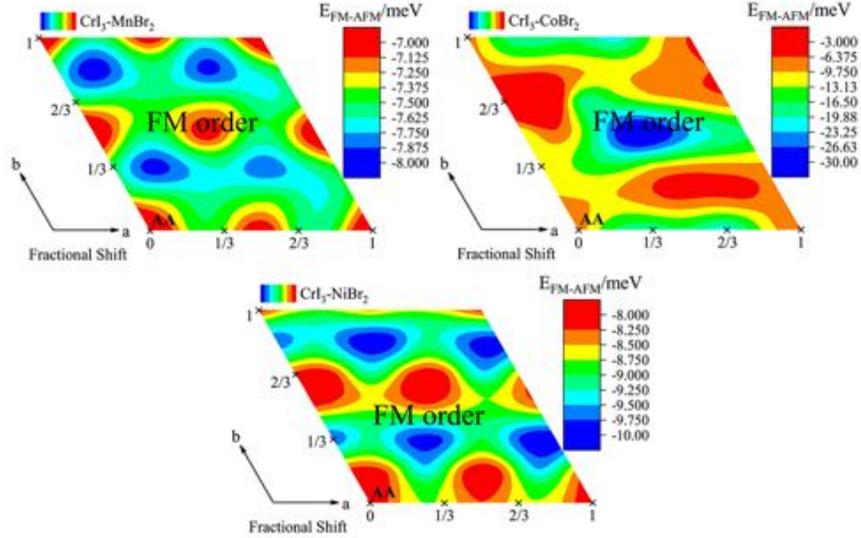

FIG. S18. The contour map for type I-II $CrI_3$-$MnBr_2$, $CrI_3$-$CoBr_2$ and $CrI_3$-$NiBr_2$ bilayers.

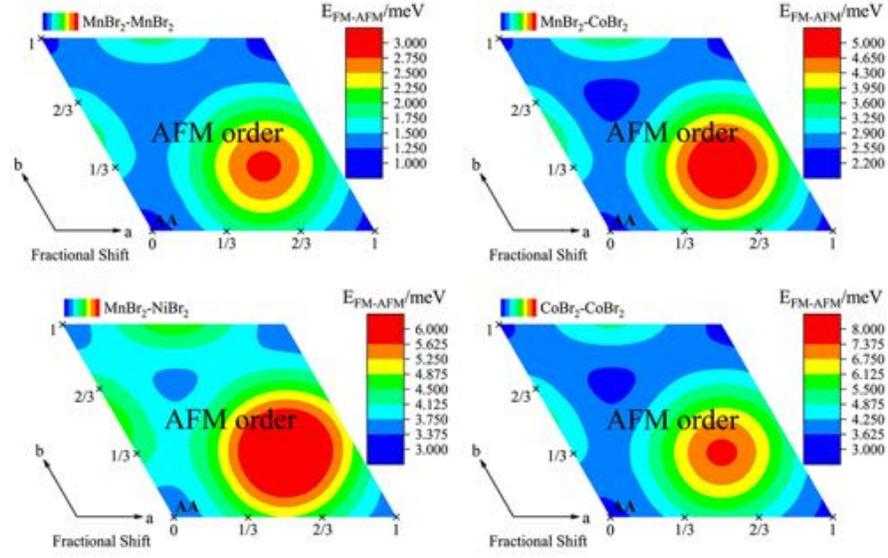

FIG. S19. The contour map for type II-II MnBr₂-MnBr₂, MnBr₂-CoBr₂, MnBr₂-NiBr₂ and CoBr₂-CoBr₂ bilayers.

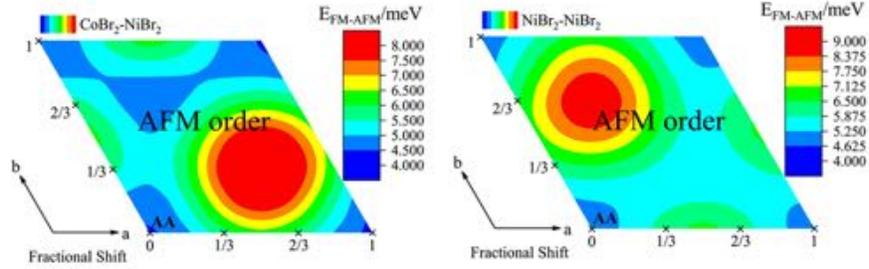

FIG. S20. The contour map for type II-II CoBr₂-NiBr₂ and NiBr₂-NiBr₂ bilayers.

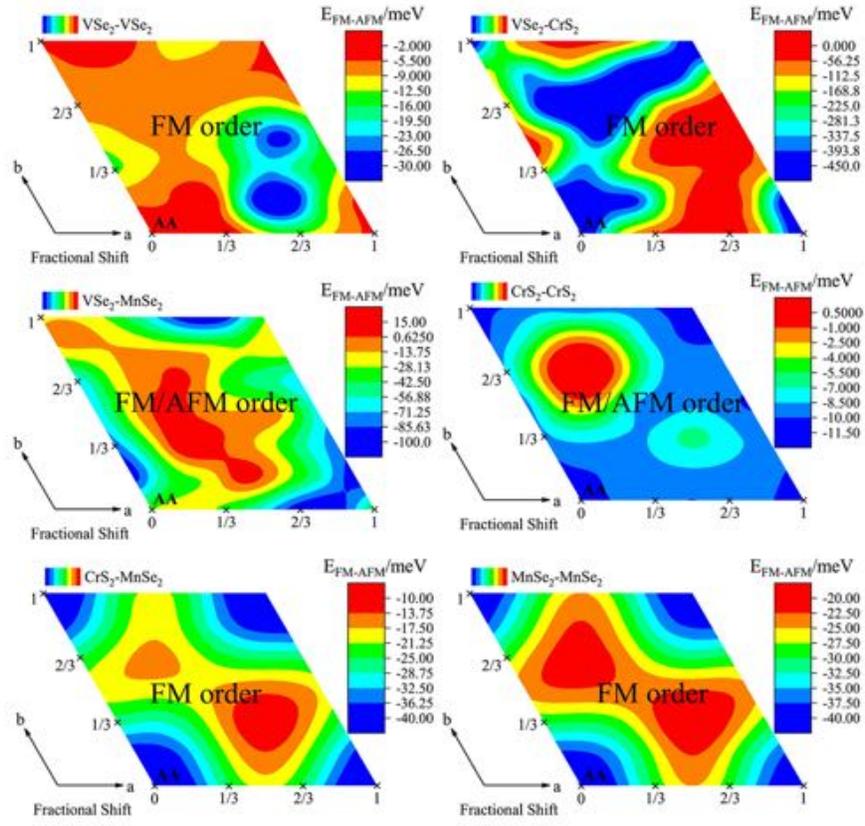

FIG. S21. The contour map for type I-I VSe$_2$-VSe$_2$, VSe$_2$-CrS$_2$, VSe$_2$-MnSe$_2$, CrS$_2$-CrS$_2$, CrS$_2$-MnSe$_2$ and MnSe$_2$-MnSe$_2$ bilayers.

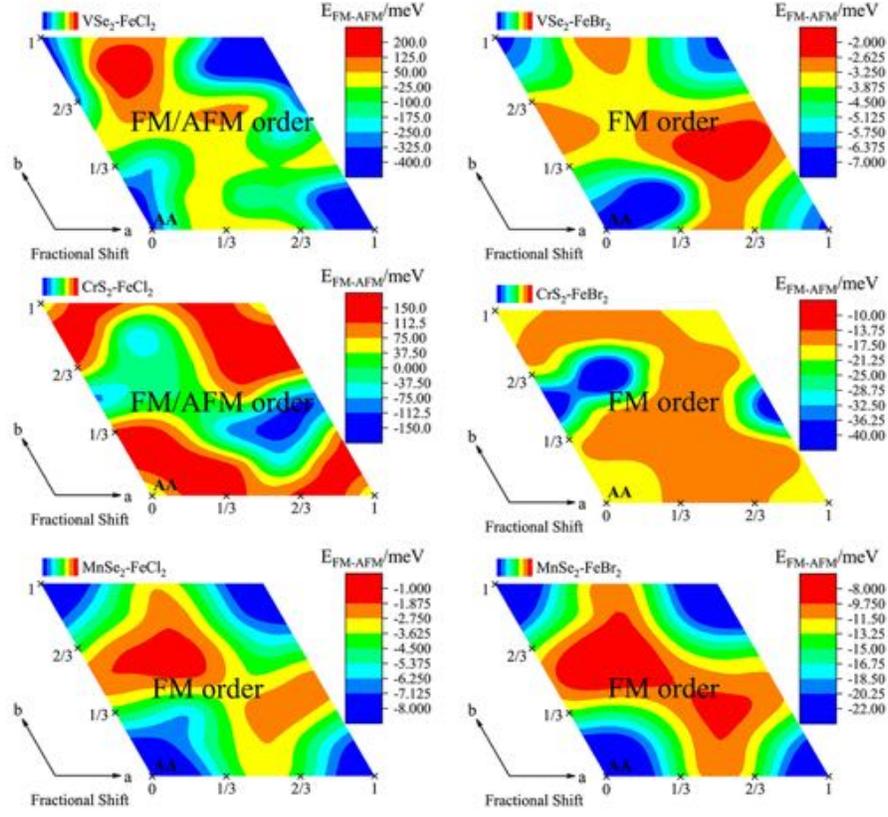

FIG. S22. The contour map for type I-II VSe$_2$-FeCl$_2$, VSe$_2$-FeBr$_2$, CrS$_2$-FeCl$_2$, CrS$_2$-FeBr$_2$, MnSe$_2$-FeCl$_2$ and MnSe$_2$-FeBr$_2$ bilayers.

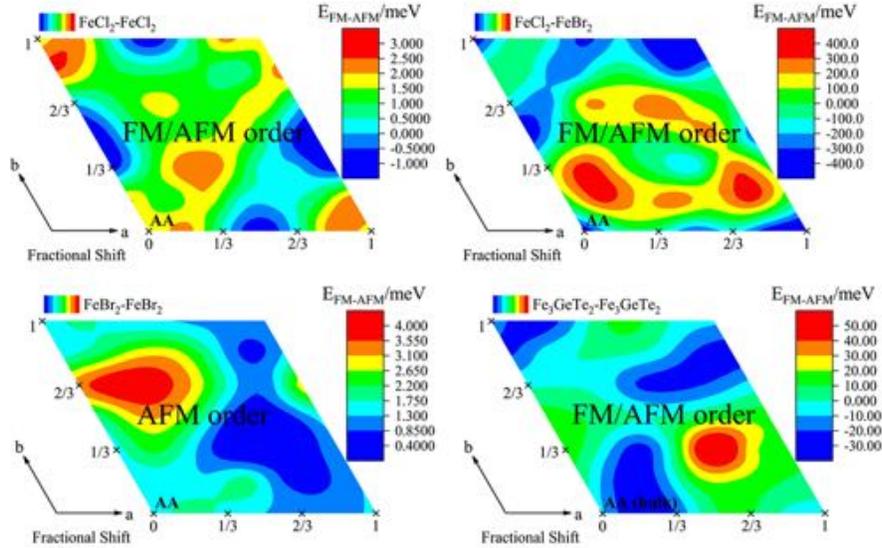

FIG. S23. The contour map for type II-II FeCl$_2$-FeCl$_2$, FeCl$_2$-FeBr$_2$, FeBr$_2$-FeBr$_2$ and Fe$_3$GeTe$_2$-Fe$_3$GeTe$_2$ bilayers.



\end{document}